\documentclass[english,british]{article}
\usepackage[T1]{fontenc}
\usepackage[latin9]{inputenc}
\usepackage[a4paper]{geometry}
\usepackage{color}
\usepackage{babel}
\usepackage{array}
\usepackage{float}
\usepackage{units}
\usepackage{amsmath}
\usepackage{graphicx}
\usepackage{esint}
\usepackage[unicode=true,pdfusetitle,
 bookmarks=true,bookmarksnumbered=false,bookmarksopen=false,
 breaklinks=false,pdfborder={0 0 1},backref=false,colorlinks=true]
 {hyperref}
\hypersetup{pdfstartview={FitH}}
\makeatletter

\providecommand{\tabularnewline}{\\}

\newcommand{\lyxaddress}[1]{
\par {\raggedright #1
\vspace{1.4em}
\noindent\par}
}

\makeatother

\begin{document}

\title{An exclusion process on a tree with constant aggregate hopping rate}

\author{Peter Mottishaw,  Bart\l{}omiej Waclaw and Martin R. Evans}

\maketitle

\lyxaddress{\noindent \begin{center}
Department of Physics and Astronomy, University of Edinburgh, Edinburgh
EH9 3JZ
\par\end{center}}
\begin{abstract}
We introduce a model of a totally asymmetric simple exclusion process
(TASEP) on a tree network where the aggregate hopping rate is constant
from level to level. With this choice for hopping rates the model
shows the same phase diagram as the one-dimensional case. The potential
applications of our model are in the area of distribution networks;
where a single large source supplies material to a large number of
small sinks via a hierarchical network. We show that mean field theory
(MFT) for our model is identical to that of the one-dimensional TASEP
and that this mean field theory is exact for the TASEP on a tree in
the limit of large branching ratio, $b$ (or equivalently large coordination
number). We then present an exact solution for the two level tree
(or star network) that allows the computation of any correlation function
and confirm how mean field results are recovered as $b\rightarrow\infty$.
As an example we compute the steady-state current as a function of
branching ratio. We present simulation results that confirm these
results and indicate that the convergence to MFT with large branching
ratio is quite rapid. 
\end{abstract}

\section{Introduction}

The totally asymmetric simple exclusion process (TASEP) consists of
hardcore particles hopping in a preferred direction on a lattice.
It has been extensively studied both as a fundamental model of non-equilibrium
systems and as a model of transport processes occurring in natural
and artificial systems (for recent reviews see \cite{blythe_nonequilibrium_2007,chou_non-equilibrium_2011}).
The one-dimensional TASEP with open boundaries is of particular interest
as a fundamental model because it exhibits a non-trivial phase diagram
consisting of 3 phases; low density, high density and maximal current.
\cite{krug_boundary-induced_1991,derrida_exact_1992,derrida_exact_1993,schutz_phase_1993}.
Moreover, exact results have been obtained for this model: the stationary
state can be determined exactly by the matrix product ansatz \cite{derrida_exact_1993,blythe_nonequilibrium_2007}
and the Bethe ansatz has been used to compute the relaxation spectrum
and to determine further dynamical transitions\cite{de_gier_bethe_2005,proeme_dynamical_2011}.
The exact results have been extended to include partial asymmetry
in the hopping \cite{sasamoto_one-dimensional_1999,blythe_exact_2000},
to compute large deviations of the density profile \cite{derrida_exact_2003},
to construct stationary states for many species of particle \cite{derrida_exact_1993-2,evans_matrix_2009}
and to analyse dynamical properties such as fluctuations and large
deviations of the current \cite{lazarescu_exact_2011}.

The main focus has been on one-dimensional systems but recently there
has been increasing interest in more complex, network versions of
the TASEP. For example applications of the model as diverse as dynamics
of molecular motors along micro-tubules, pedestrian traffic flow and
queueing systems require transport along connected pathways. Generally,
as exact results are not available, mean-field approaches \cite{derrida_exact_1992}
have been used to study connected multi-lane systems and to predict
phase diagrams \cite{evans_phase_2011}. There has also been detailed
work on random networks of connected one-dimensional TASEPs \cite{neri_totally_2011},
again using the mean field approximation. The analysis shows that
networks introduce some interesting behaviours and the work has recently
been extended to model active motor protein transport on the cytoskeleton
\cite{neri_modelling_2012}. 

Another approach by Basu and Mohanty \cite{basu_asymmetric_2010}
is to look at a simple extension of a TASEP to a Cayley tree. This
model is interesting because it has potential application in natural
and artificial processes that take place on branching structures and
is potentially simple enough to be tractable in steady state. Basu
and Mohanty used a mean field approximation and simulations to show
that the behaviour of the model was rather straightforward. There
is only a single low density phase and effectively the particles flow
freely from the root of the tree and then wait to exit at the final
layer. The model does not show the rich behaviour of the one-dimensional
TASEP.

The model we address here is similar to the Cayley tree model, but
crucially the bulk hopping rates in our ``TASEP on a tree'' model
are defined so that the aggregate hopping rate from one level of the
tree to the next remains constant. As a result we see in steady state
a rich phase diagram and behaviour similar to the one-dimensional
case. The constant aggregate hopping rate can be achieved by having
a fixed branching ratio $b$ for the tree while at the same time reducing
the hopping rate by a factor $\frac{1}{b}$ as we move from one level
of the tree to the next starting from the root. This can be viewed
as successively dividing fixed transport capacity between an increasing
number of paths. We believe the model is more relevant in real-world
applications because it does not suffer from the exponential growth
in capacity of typical Bethe lattice models of statistical physics
which are effectively infinite dimensional.

Another important outcome from our TASEP on a tree model is that we
can obtain some exact results. The most interesting of these is to
show that the mean field theory of the one-dimensional model is actually
exact for the TASEP on a tree in the limit of large branching rate
(or equivalently large coordination number). The TASEP on a tree model
in the limit of large branching ratio is a simpler model than the
one-dimensional TASEP but has a similarly rich phase diagram. It has
the potential to play a similar role to the infinite dimensional models
in equilibrium statistical mechanics. Another exact result is the
full solution of the two level tree or star network.

The potential applications of our model are in the area of distribution
networks; where a single large source supplies material to a large
number of small sinks via a hierarchical network. This type of network
has been used to model a variety of natural and artificial systems,
such as the behaviour of river systems, arterial blood flow and city
traffic; see for example \cite{banavar_general_2010,dodds_optimal_2010}
and references therein. The optimal design of these systems has been
the main focus of this research and it has been assumed that the steady-state
current in the system is determined solely by the capacity of the
distribution network. The model we analyse here attempts to answer
a different question; what steady-state current is achieved for specific
values of the source input rate, transport rate (in the distribution
network) and sink output rate. Our results show that the steady-state
current has a non-trivial dependence on these parameters. Interestingly,
the two level tree, star network or explosion network is the optimal
design under certain assumptions \cite{gastner_spatial_2006,dodds_optimal_2010,banavar_general_2010}
and for this we provide a full analytic solution. It is an open question
as to the implications of our results in the different application
areas, but they are most relevant to applications, such as vehicular
traffic, where the material being distributed can be modelled by the
hopping of hardcore particles.

The outline of this paper is as follows. In Section \ref{sec:Model-definition}
we define the TASEP on a tree model and use simple arguments to show
that mean field theory (MFT) for the tree is identical to the one-dimensional
model. In Section \ref{sec:Dynamics-and-the} we derive the steady-state
equations from the master equation and confirm the MFT results. In
Section \ref{sec:Correlation-functions} we extend the master equation
approach to arbitrary correlation functions and show that MFT is exact
in the limit of large coordination number and when the total boundary
hopping rates sum to unity. In Section \ref{sec:Exact-solution-for}
we provide a detailed analysis of the exact solution of the two level
tree. Finally in Section \ref{sec:Simulation-results} we provide
detailed results for simulations of the TASEP on a tree on finite
latices and compare them with MFT and exact results for the one-dimensional
model.

\section{\label{sec:Model-definition}Model definition and mean field theory }

\subsection{Model definition }

The TASEP in one dimension consists of particles hopping in one direction
along the lattice. Each site can be occupied by at most one particle.
Hopping is only allowed from a given site to its next neighbour to
the right and a particle is blocked from hopping if the neighbouring
site is occupied. In the model with open boundaries particles are
injected at the first site at rate $\alpha$ and removed at the last
site at rate $\beta$. In the simplest case the bulk hopping rate
between sites is the same for all sites and can be taken as $1$ without
loss of generality. In this paper we will generalise the one-dimensional
model to a tree lattice while retaining these interesting characteristics.

\begin{figure}
\noindent \begin{centering}
\includegraphics[scale=0.8]{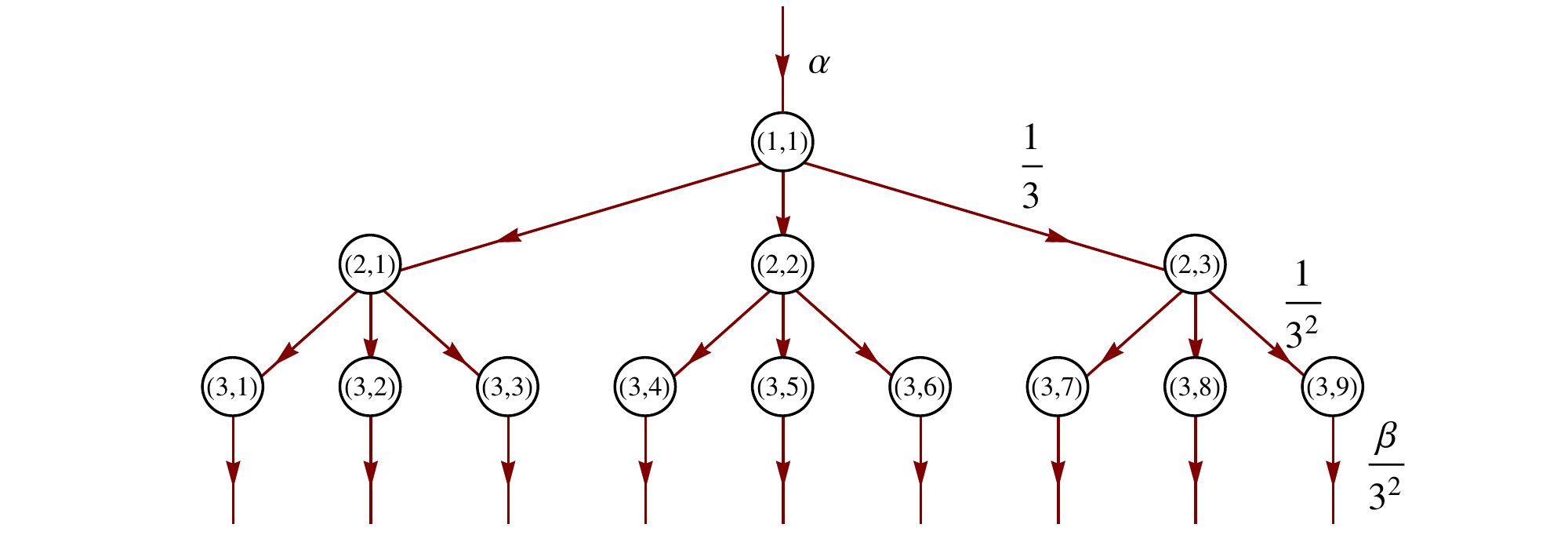}
\par\end{centering}

\caption{\label{fig:tree}TASEP on a tree with $K=3$ and $b=3$. The notation
for labelling sites is $\left(k,i\right)$; where $k$ labels the
level in the tree and $i$ labels the site at that level. The number
of sites at level $k$ is $b^{k-1}$. The hopping rates at each level
are given on the right. The hopping rates are scaled so that the ``total''
hopping rate from one level to the next is unity and independent of
$b$. The exit hopping rates at the final level are similarly scaled
so that the ``total'' exit rate is $\beta$ independent of $b$.
These ``total'' hopping rates are therefore identical to the one-dimensional
case.}
\end{figure}

We generalise the one-dimensional case by introducing a branching
number $b$. The tree lattice can be constructed by starting with
a single ``root'' site where particles are injected. This site is
labelled $\left(1,1\right)$, indicating level 1 of the tree and site
1 at that level. Site $\left(1,1\right)$ is connected to $b$ level
2 sites labelled $\left(2,1\right),$$\left(2,2\right),\ldots,\left(2,b\right)$
where the first number indicates these sites are at level 2. Each
of these are given $b$ neighbours at level 3. The process is repeated
up to level $K$ where particles can exit from the lattice. The notation
we use is that each site is labelled by a pair of indices $\left(k,i\right)$,
where $k$ is the level in the tree ($k=1,2,3,\ldots,K$) and $i$
is a site index for that level ($i=1,2,3,\ldots,b^{k-1}$). An example
of this structure is shown in figure \ref{fig:tree}. Each site $\left(k,i\right)$
can be either empty ($\tau_{k,i}=0$) or occupied by a single particle
($\tau_{k,i}=1$). Hopping is only allowed from a given site to one
if its vacant neighbours at the next level. Clearly with $b=1$ this
reduces to a one-dimensional TASEP model.

Basu and Mohanty \cite{basu_asymmetric_2010} define hopping rates
that significantly reduce the exclusion effect and lead to a ``free
flow'' of particles to the exit boundary and rather straightforward
behaviour. Here we take an alternative approach where the hopping
rate is reduced by a factor of $\frac{1}{b}$ at each level, so that
the hopping rate from a site at level $k$ to a connected site at
level $k+1$ is $1/b^{k}$. The number of sites increases by a factor
of $b$ at each level so that the ``overall'' hopping rate from
one level to the next remains of the same order. At the exit boundary
(level $K$) there are $b^{K-1}$ sites labelled $\left(K,i\right)$
with $i=1,\ldots,b^{K-1}$ . We give each site an exit rate $\beta/b^{K-1}$
so that the ``overall'' exit rate is of order $\beta$ . With this
definition of the TASEP on a tree the overall entry rate is $\alpha$
and the overall removal rate is $\beta$ as in the normal one-dimensional
TASEP. With $b=1$ the hopping rates reduces to the usual open TASEP
model.

The dynamics can be summarised as follows. During every infinitesimal
time interval $\mathrm{d}t$, each particle at a site $\left(k,i\right)$
with $1\leq k\leq K-1$ (i.e. bulk sites with $\tau_{k,i}=1$) will
hop to an empty connected neighbour at level $k+1$ (i.e. neighbours
with $\tau_{k+1,j}=0$) with probability $\mathrm{d}t/b^{k-1}$. During
the same infinitesimal time interval, if site $\left(1,1\right)$
is empty ($\tau_{1,1}=0$) a particle is added with probability $\alpha\mathrm{d}t$
and any particle at level $K$ is removed with probability $\beta\mathrm{d}t/b^{K-1}$.

\subsection{Steady-state current}

The current of particles $j_{k,i}$ into a given site $\left(k,i\right)$
depends on the probability that the site is vacant, its downstream
neighbour is occupied and the hopping probability. This leads to expressions
for the average current into site $\left(k,i\right)$ in the bulk
and at the two boundaries
\begin{eqnarray}
j_{1,1} & = & \alpha\left(1-\left\langle \tau_{1,1}\right\rangle \right)\label{eq:j1 current}\\
j_{k,i} & = & \frac{1}{b^{k-1}}\left\langle \tau_{k-1,\left\lceil \nicefrac{i}{b}\right\rceil }\left(1-\tau_{k,i}\right)\right\rangle \label{eq:jk current}\\
j_{K+1,i} & = & \frac{\beta}{b^{K-1}}\left\langle \tau_{K,i}\right\rangle \label{eq:jK current}
\end{eqnarray}

where $\left\lceil x\right\rceil $ is the ceiling function defined
to be the smallest integer greater than or equal to $x$. The currents
are time-dependent and the average is over a time-dependent probability
distribution for the $\tau_{k,i}$ which is dependent on initial conditions.
We are interested in the steady state where the system has evolved
for long enough that all correlation functions are independent of
time and all memory of initial conditions has been lost. In particular
any asymmetry in the initial conditions will have been lost and from
the symmetry of the lattice we have in steady state
\begin{equation}
\left\langle \tau_{k,i}\left(1-\tau_{k+1,bi-b+j}\right)\right\rangle \textnormal{is independent of}\, j\,\textnormal{for}\,1\leq j\leq b\textnormal{.}\label{eq:symmetry}
\end{equation}

If we define the total current $J_{k}$ flowing into level $k$ as
$J_{k}=\sum_{i}j_{k,i}$ then in steady state we have
\begin{eqnarray}
J_{1} & = & \alpha\left(1-\left\langle \tau_{1}\right\rangle \right)\label{eq:J1 total current}\\
J_{k} & = & \left\langle \tau_{k-1}\left(1-\tau_{k}\right)\right\rangle \label{eq:Jk total current}\\
J_{K+1} & = & \beta\left\langle \tau_{K}\right\rangle \label{eq:JK total current}
\end{eqnarray}

where all the two-point correlation functions are between connected
sites on adjacent levels in the tree and we have suppressed the site
index $i$ on the $\tau_{k,i}$ because in steady state the average
occupation number and the connected two point correlation function
depend only on the level index, $k$. In steady state the average
site occupancy is independent of time and therefore we expect the
total current into each level to be the same steady-state current,
$J=J_{k}$, independent of $k$. We can therefore obtain a set of
equations for the expectation values by equating the right hand side
of each of equation (\ref{eq:J1 total current}), (\ref{eq:Jk total current})and
(\ref{eq:JK total current}). There is no explicit dependence on $b$
in these equations suggesting that the behaviour of the tree model
will be the same as the one-dimensional case. However, as we show
later, the higher level correlation functions do have an explicit
dependence on $b$ leading to different behaviour from the one-dimensional
model. The steady-state current equations are insufficient to determine
the $\left\langle \tau_{k}\right\rangle $ without some approximation
because of the coupling to higher level correlation functions.

\subsection{Mean field theory\label{sub:Mean-field-theory} }

In the mean field approximation \cite{derrida_exact_1992} we ignore
correlations between sites and in particular take $\left\langle \tau_{k-1}\left(1-\tau_{k}\right)\right\rangle =t_{k-1}\left(1-t_{k}\right)$
where $t_{k}=\left\langle \tau_{k}\right\rangle $. If we substitute
these in the total current equations (\ref{eq:J1 total current}),
(\ref{eq:Jk total current}) and (\ref{eq:JK total current}), and
use the steady-state condition that current is constant we obtain
the mean field theory equations
\begin{eqnarray}
\alpha\left(1-t_{1}\right) & = & t_{1}\left(1-t_{2}\right)\label{eq:mft_alpha}\\
t_{k-1}\left(1-t_{k}\right) & = & t_{k}\left(1-t_{k+1}\right)\label{eq:mft_bulk}\\
t_{K-1}\left(1-t_{K}\right) & = & \beta t_{k}\,.\label{eq:mft_beta}
\end{eqnarray}

\begin{figure}
\noindent \begin{centering}
\includegraphics[scale=0.8]{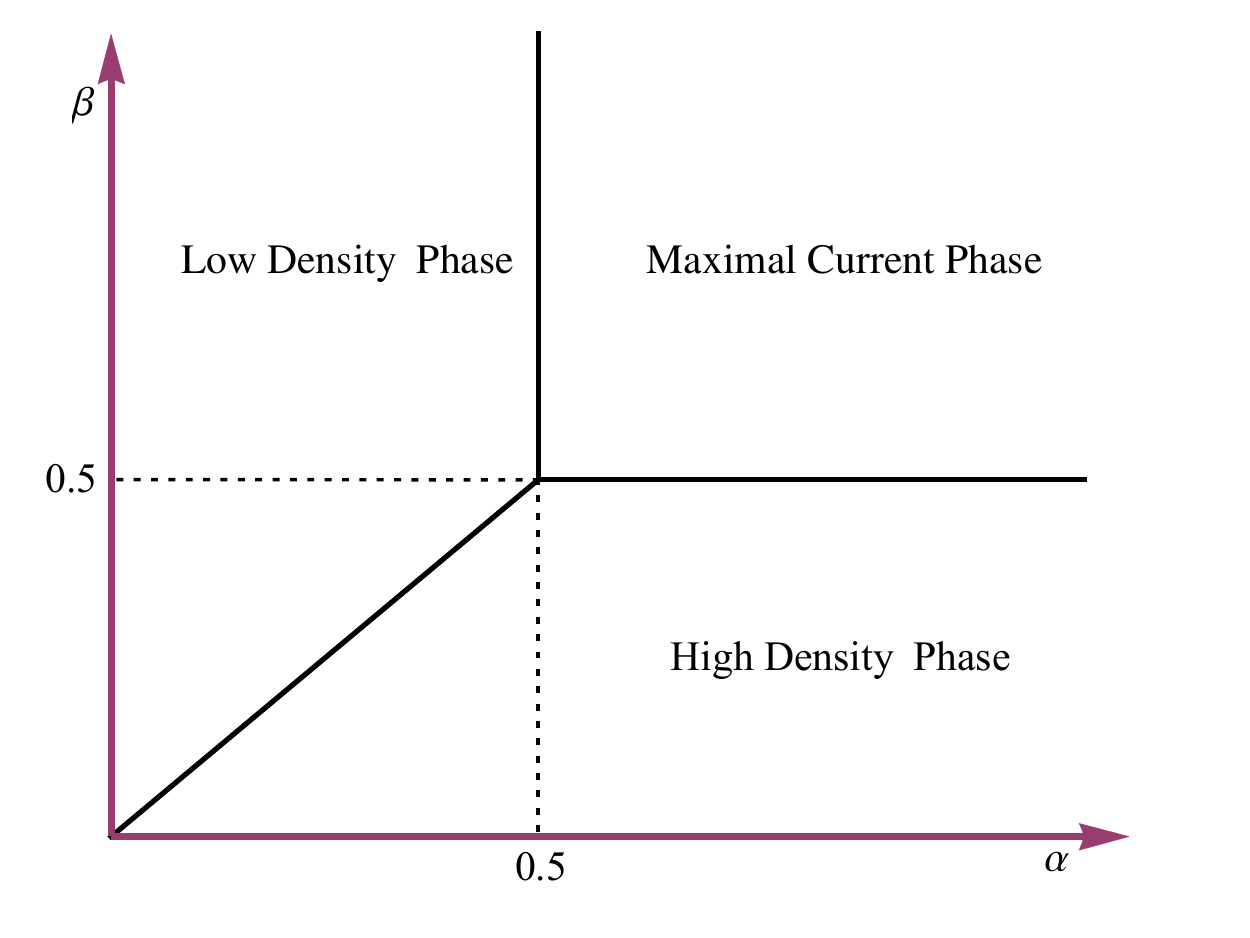}\caption{\label{fig:Phase-diagram}Mean field phase diagram for TASEP on a
tree.}

\par\end{centering}

\end{figure}

These are independent of $b$ so that at the level of mean field theory
the model defined on the tree is formally identical to the one-dimensional
case. Therefore at the mean field theory level all tree models considered
in this paper will have the same phase diagram as the one-dimensional
MFT phase diagram (see \cite{derrida_exact_1992}) shown in figure
\ref{fig:Phase-diagram}. However, $t_{k}$ is the average occupancy
of a site at level $k$ not the average occupancy of the level overall
which grows exponentially. 

As in the one-dimensional case there are three phases determined by
the boundary hopping rates $\alpha$ and $\beta$. These are most
easily understood in terms of the iteration of the MFT equations(\ref{eq:mft_alpha})(\ref{eq:mft_bulk})
and (\ref{eq:mft_beta}) (see \cite{derrida_exact_1992}). When the
steady-state current $J<\frac{1}{4}$ there is a high density stable
fixed point and a low density unstable fixed point with densities
given by
\[
t_{\pm}=\frac{1}{2}\left[1\pm\sqrt{1-4J}\right].
\]

In the high density phase $\alpha>\beta$ and $\alpha<\frac{1}{2}$.
The density rapidly converges to the stable high density fixed point
and the current is given by $J=\beta\left(1-\beta\right)$. In the
low density phase $\alpha<\beta$ and $\beta<\frac{1}{2}$. The initial
density $t_{1}$ is set infinitesimally close to the unstable low
density fixed point and only diverges from it as the exit boundary
is approached. The steady-state current is given by $J=\alpha\left(1-\alpha\right)$. 

In the maximal current phase $\alpha>\frac{1}{2}$ and $\beta>\frac{1}{2}$.
The current is $J=\frac{1}{4}$ and there is a single marginal fixed
point (there are no fixed points for $J>\frac{1}{4}$). The initial
density $t_{1}$ is set above the marginal fixed point. The density
rapidly approaches the fixed point but moves below it as the exit
boundary is approached.

We have seen that at the mean field level the tree model is the same
as the one-dimensional model. In the rest of the paper we will go
beyond MFT and explore correlations. This will allow us to provide
some results on the $b$ dependence of the correlation functions.

\section{\label{sec:Dynamics-and-the}Dynamics and the master equation }

As a next step to understanding the behaviour of the tree model we
derive the exact equations of motion for the average occupation number,
$\left\langle \tau_{k,i}\right\rangle $ and then look at the steady-state
behaviour. This approach enables us to recover the results of Section
\ref{sec:Model-definition} but has the advantage of being applicable
to an arbitrary correlation function as we will show in Section \ref{sec:Correlation-functions}.

\subsection{Continuous time dynamics and the master equation}

As a starting point for obtaining the equations of motion we use a
master equation to describe the evolution of the the probability $P\left(\mathcal{C},t\right)$
of the system being in configuration $\mathcal{C}\equiv\left\{ \tau_{k,i}\right\} $
at time $t$ given some set of initial conditions
\begin{equation}
\frac{\partial P\left(\mathcal{C},t\right)}{\partial t}=\sum_{\mathcal{C}^{\prime}}P\left(\mathcal{C}^{\prime},t\right)W\left(\mathcal{C}^{\prime}\rightarrow\mathcal{C}\right)-\sum_{\mathcal{C}^{\prime\prime}}P\left(\mathcal{C},t\right)W\left(\mathcal{C}\rightarrow\mathcal{C}^{\prime\prime}\right)\label{eq:master-simple}
\end{equation}

where the transition rates are defined by
\begin{equation}
W\left(\mathcal{C^{\prime}}\rightarrow\mathcal{C}\right)=\begin{cases}
\alpha & \textrm{if }\mathcal{C}\,\textnormal{is identical to }\mathcal{C^{\prime}\,}\textnormal{except for an additonal particle at}\left(1,1\right)\textnormal{ }\\
\nicefrac{\beta}{b^{K-1}} & \textnormal{if \ensuremath{\mathcal{C^{\prime}\,}}is identical to \ensuremath{\mathcal{C}}\,}\textnormal{except for an additional particle at level }K\\
\nicefrac{1}{b^{k}} & \textnormal{if }\mathcal{C\,}\textnormal{and }\mathcal{C^{\prime}\,}\textnormal{differ on a single pair of connected sites }\left(k,i\right)\\
 & \textnormal{and any one of }\left(k+1,bi-b+1\right),\ldots,\left(k+1,bi\right)\\
0 & \textnormal{otherwise.}
\end{cases}\label{eq:Hopping rates}
\end{equation}

With this definition for the transition rates we have precisely the
dynamics we defined in section \ref{sec:Model-definition}.

\subsection{Exact equations for the time evolution of the average site density}

The equation of motion for the expectation value of the occupation
number, $\left\langle \tau_{k,i}\right\rangle $, is given by
\[
\frac{\partial\left\langle \tau_{k,i}\right\rangle }{\partial t}=\sum_{\mathcal{C}^{\prime}}P\left(\mathcal{C}^{\prime},t\right)\sum_{C\neq C^{\prime}}\tau_{k,i}W\left(\mathcal{C}^{\prime}\rightarrow\mathcal{C}\right)-\sum_{\mathcal{C}}P\left(\mathcal{C},t\right)\tau_{k,i}\sum_{C^{\prime\prime}}W\left(\mathcal{C}\rightarrow\mathcal{C}^{\prime\prime}\right)
\]

where we have used the master equation (\ref{eq:master-simple}) and
taken $\mathcal{C}\equiv\left\{ \tau_{k,i}\right\} $ , so that configuration
$\mathcal{C}$ determines the values of $\tau_{k,i}$. Substituting
for the transition rates from equation (\ref{eq:Hopping rates}) gives
three equations for the time evolution of the average occupation number
in terms of two-point correlation functions
\begin{eqnarray}
\frac{\partial\left\langle \tau_{1,1}\right\rangle }{\partial t} & = & \alpha\left(1-\left\langle \tau_{1,1}\right\rangle \right)-\frac{1}{b}\sum_{i=1}^{b}\left\langle \tau_{1,1}\left(1-\tau_{2,i}\right)\right\rangle \label{eq:11_eq_of_motion}\\
\frac{\partial\left\langle \tau_{k,i}\right\rangle }{\partial t} & = & \frac{1}{b^{K-1}}\left\langle \tau_{k-1,\left\lceil \nicefrac{i}{b}\right\rceil }\left(1-\tau_{k,i}\right)\right\rangle -\frac{1}{b^{k}}\sum_{j=0}^{b-1}\left\langle \tau_{k,i}\left(1-\tau_{k+1,bi-j}\right)\right\rangle \label{eq:ki_eq_of_motion}\\
\frac{\partial\left\langle \tau_{K,j}\right\rangle }{\partial t} & = & \frac{1}{b^{K-1}}\left\langle \tau_{K-1,\left\lceil \nicefrac{j}{b}\right\rceil }\left(1-\tau_{K,j}\right)\right\rangle -\frac{\beta}{b^{K-1}}\left\langle \tau_{K,j}\right\rangle \label{eq:Kj_eq_of_motion}
\end{eqnarray}

where $1\leq i\leq b^{k-1}$ in the second equation and $1\leq j\leq b^{K-1}$
in the third equation.

\subsection{Steady-state equations of motion}

In steady state any asymmetry in the initial conditions will have
been lost and from the symmetry of the lattice the steady-state form
of equations (\ref{eq:11_eq_of_motion}), (\ref{eq:ki_eq_of_motion})
and (\ref{eq:Kj_eq_of_motion}) is
\begin{eqnarray}
0 & = & \alpha\left(1-\left\langle \tau_{1}\right\rangle \right)-\left\langle \tau_{1}\left(1-\tau_{2}\right)\right\rangle \label{eq:1_correlation}\\
0 & = & \left\langle \tau_{k-1}\left(1-\tau_{k}\right)\right\rangle -\left\langle \tau_{k}\left(1-\tau_{k+1}\right)\right\rangle \label{eq:k_correlation}\\
0 & = & \left\langle \tau_{K-1}\left(1-\tau_{K}\right)\right\rangle -\beta\left\langle \tau_{K}\right\rangle \label{eq:K_correlation}
\end{eqnarray}

where all the two-point correlation functions are between connected
sites on adjacent levels in the tree and we have suppressed the site
index. The lattice symmetry that is used here is that the steady-state
equations are invariant under transformations of the lattice that
preserve the connectivity of the tree. For example, any two sub branches
that have their ``root'' at the same site can be interchanged without
changing the steady-state equations.

These are identical to the total steady-state current equations (\ref{eq:J1 total current}),
(\ref{eq:Jk total current}) and (\ref{eq:JK total current}), obtained
using the assumption of constant current. They are also the same as
the one-dimensional TASEP equations. However, the master equation
approach used in this section is more easily extended to higher-order
correlation functions.

\section{\label{sec:Correlation-functions}Correlation functions }

We now consider the equation of motion for a general correlation function.
In the case of a TASEP where $\tau_{i}$ can only take the values
one or zero the most general correlation function is
\begin{equation}
c\left(S,t\right)\equiv\left\langle \prod_{i\in S}\tau_{i}\right\rangle \equiv\sum_{\mathcal{C}}P\left(\mathcal{C},t\right)\prod_{i\in S}\tau_{i},\label{eq:crln_fncn_dfnn}
\end{equation}

where $S$ is any given subset of sites in the lattice. In appendix
\ref{sec:Equations-for-time} we derive the equation of motion for
$c\left(S,t\right)$ for the TASEP on an arbitrary lattice. In this
section we will use this for correlation functions on the tree.

\subsection{Time evolution equation for tree lattice }

We would like to generalise the steady-state equations (\ref{eq:1_correlation}),
(\ref{eq:k_correlation}) and (\ref{eq:K_correlation}) to the general
correlation function (\ref{eq:crln_fncn_dfnn}). In appendix \ref{sec:Equations-for-time}
we consider a general open TASEP on an arbitrary network and obtain
the equation of motion for the correlation functions; equation (\ref{eq:general_eqtn_of_mtn})
. The steady-state form of this for a subset of sites $S$ on a tree
is
\begin{eqnarray}
0 & = & \alpha\left\{ c\left(S\left[\overline{\left(1,1\right)}\right]\right)-c(S)\right\} \delta\left[\left(1,1\right)\in S\right]\nonumber \\
 & + & \sum_{\left(i,j\right)\in S_{in}}\frac{1}{b^{i-1}}\left\{ c\left(S\left[\left(i-1,\left\lceil \nicefrac{j}{b}\right\rceil \right),\overline{\left(i,j\right)}\right]\right)-c\left(S\left[\left(i-1,\left\lceil \nicefrac{j}{b}\right\rceil \right)\right]\right)\right\} \nonumber \\
 & - & \sum_{\left(i,j\right)\in S_{out}}\frac{1}{b^{i}}\sum_{(i+1,k)\in O\left(i,j\right)}\left\{ c(S)-c\left(S\left[\left(i+1,k\right)\right]\right)\right\} \nonumber \\
 & - & \frac{\beta}{b^{K-1}}\sum_{i\in S_{exit}}c(S)\label{eq:bl_steadystate}
\end{eqnarray}

where we have discarded the $t$ dependence of the correlation function in 
 (\ref{eq:crln_fncn_dfnn}) and the $\delta[\textrm{statement}]$ is a generalised Kronecker
delta (or indicator function); equal to unity if statement is true
and zero otherwise. The notation $S\left[\overline{\left(1,1\right)}\right]$
means the subset of sites obtained by
removing site $\left(1,1\right)$ from the set of sites $S$. Similarly,
$S\left[\left(i,k\right)\right]$ means the subset of sites obtained by adding site $\left(i,k\right)$
to $S$ and $S\left[\left(i,k\right),\overline{\left(j,l\right)}\right]$
means the subset of sites obtained by
adding site $\left(i,k\right)$ to $S$ and removing site $\left(j,l\right)$.
The various subsets of $S$ ($S_{entry}$, $S_{exit}$, $S_{in}$,
$S_{out}$) that appear in the sums are defined in table \ref{tab:Definition-of-boundary}.
The outgoing neighbour set $O\left(i\right)$ and the incoming neighbour
set $I\left(j\right)$ are also defined in table \ref{tab:Definition-of-boundary}. 

\begin{table}

\noindent \centering{}%
\begin{tabular}{|c||>{\raggedright}p{12cm}|}
\hline 
\textbf{Subsets} & \textbf{Definition of boundary sites related to $S$.}\tabularnewline
\hline 
\hline 
$S_{entry}$ & The subset of $S$ at which particles can enter $S$ from outside
the lattice.\tabularnewline
\hline 
\hline 
$S_{exit}$ & The subset of $S$ at which particles can leave $S$ and exit the
lattice.\tabularnewline
\hline 
\hline 
$S_{in}$ & The subset of $S$ where particles can enter $S$ from lattice sites
that are not in $S$.\tabularnewline
\hline 
\hline 
$S_{out}$ & The subset of $S$ where particles can leave $S$ to sites in the
lattice that are not in $S$.\tabularnewline
\hline 
\hline 
 $O\left(i\right)$ & The set of outgoing neighbours to site $i\in S_{out}$ that are not
in $S$.\tabularnewline
\hline 
\hline 
$I\left(j\right)$ & The set of incoming neighbours to site $j\in S_{in}$ that are not
in $S$.\tabularnewline
\hline 
\end{tabular}
\caption{\label{tab:Definition-of-boundary}Definition of subsets of boundary sites related
to $S$}
\end{table}

We see that the steady-state equations (\ref{eq:1_correlation}),
(\ref{eq:k_correlation}) and (\ref{eq:K_correlation}) that relate
the single-point correlation functions to the two-point correlation
functions are the first in a hierarchy of equations. However, the
branching ratio $b$ drops out of the lowest order equation leading
to a mean field theory that is independent of $b.$ We see that this
is not the case for other levels in the hierarchy and so any corrections
to mean field theory should be dependent on $b$.

\subsection{Exact solution when $\alpha+\beta=1$ }

An interesting question is whether mean field theory is exact in some
situations. This would require a solution of the form
\begin{equation}
c(S)=\prod_{\left(k,j\right)\in S}t_{k}\label{eq:mf_ansatz}
\end{equation}

for all possible $S$ where $t_{k}$ satisfies the mean field equations
(\ref{eq:mft_alpha}), (\ref{eq:mft_bulk}) and (\ref{eq:mft_beta}).
It is easy to show that where $S$ contains just a single site this
equation is satisfied on the line $\beta=1-\alpha$ with $t_{k}=\alpha$,
but we need to demonstrate this for any $S$. We take $\beta=1-\alpha$
and substitute the ansatz
\begin{equation}
c(S)=\alpha^{\left|S\right|}\label{eq:mf_solution}
\end{equation}

into equation (\ref{eq:bl_steadystate}). This becomes
\begin{eqnarray}
0 & = & \alpha^{\left|S\right|}\left(1-\alpha\right)\delta\left[\left(1,1\right)\in S\right]\nonumber \\
 & + & \alpha^{\left|S\right|}\left(1-\alpha\right)\sum_{\left(i,j\right)\in S_{in}}\frac{1}{b^{i-1}}\nonumber \\
 & - & \alpha^{\left|S\right|}\left(1-\alpha\right)\sum_{\left(i,j\right)\in S_{out}}\frac{1}{b^{i}}\sum_{(i+1,k)\in O\left(i,j\right)}\nonumber \\
 & - & \frac{\alpha^{\left|S\right|}\left(1-\alpha\right)}{b^{K-1}}\sum_{i\in S_{exit}}.\label{eq:bl_steadystate-1}
\end{eqnarray}

We can eliminate the common factor to obtain
\begin{eqnarray}
0 & = & \delta\left[\left(1,1\right)\in S\right]+\sum_{\left(i,j\right)\in S_{in}}\frac{1}{b^{i-1}}-\sum_{\left(i,j\right)\in S_{out}}\frac{1}{b^{i}}\sum_{(i+1,k)\in O\left(i,j\right)}1-\sum_{i\in S_{exit}}1.\label{eq:line_of_mean_field}
\end{eqnarray}

This depends only on the properties of the lattice and not on the
value of $\alpha$. We provide an inductive proof that it is satisfied
for any $S$ in appendix \ref{sec:Inductive-proof-that}. This result
is independent of $b$ and shows that the one-dimensional result \cite{derrida_exact_1992}
generalises to the tree.

\subsection{\label{sub:Expansion-of-correlation}Expansion of correlation functions
in $1/b$ and mean field theory }

There is another more general situation where mean field theory is
exact; this is in the limit $b\rightarrow\infty$. First we shall
give a simple heuristic argument for the result and then outline the
proof. There are no closed loops in the tree so that in general neighbouring
sites are correlated because the probability of hopping from a site
at level $k$ is reduced if its neighbours at level $k+1$ are occupied.
However, for a given site, as the number of these downstream nearest
neighbour sites increases the effect of any single downstream nearest
neighbour site will diminish and we expect that as $b\rightarrow\infty$
this mechanism eliminates any correlation between pairs of connected
sites. We can put this plausibility argument on a more rigorous footing
by considering an expansion of the correlation function in terms of
$1/b$.

The correlation functions that we defined as
\begin{equation}
c\left(S\right)\equiv\left\langle \prod_{\left(k,i\right)\in S}\tau_{k,i}\right\rangle 
\end{equation}

are dependent on $b,$ so we can expand $c\left(S\right)$ as a power
series in $1/b$
\begin{equation}
c\left(S\right)=\sum_{r=0}^{\infty}c_{r}\left(S\right)\frac{1}{b^{r}}\label{eq:large_b_expansion}
\end{equation}

where the coefficients $c_{r}\left(S\right)$ are independent of $b$.
Our hypothesis is that for a fixed choice of $S$, as $\frac{1}{b}\rightarrow0$,
the correlation function should satisfy mean field theory. This requires
the zeroth order term to satisfy
\begin{equation}
c_{0}\left(S\right)=\prod_{k=1}^{K}t_{k}^{\left|S\left(k\right)\right|}\label{eq:mft_ansatz}
\end{equation}

where $\left|S\left(k\right)\right|$ is the number of sites in $S$
at level $k$ of the tree and the $t_{k}$ satisfy the mean field
equations (\ref{eq:mft_alpha}), (\ref{eq:mft_bulk}) and (\ref{eq:mft_beta}).
In order to obtain this result we substitute the expansion (\ref{eq:large_b_expansion})
in (\ref{eq:bl_steadystate}). Retaining only the lowest order terms
in $1/b$ gives
\begin{eqnarray}
0 & = & \alpha\left|S\left(1\right)\right|\left\{ c_{0}\left(S\left[\overline{\left(1,1\right)}\right]\right)-c_{0}\left(S\right)\right\} \nonumber \\
 & + & \sum_{\left(i,j\right)\in S_{in}}\frac{1}{b^{i-1}}\left\{ c_{0}\left(S\left[\left(i-1,\left\lceil \nicefrac{j}{b}\right\rceil \right),\overline{\left(i,j\right)}\right]\right)-c_{0}\left(S\left[\left(i-1,\left\lceil \nicefrac{j}{b}\right\rceil \right)\right]\right)\right\} \nonumber \\
 & - & \sum_{\left(i,j\right)\in S_{out}}\frac{1}{b^{i}}\sum_{(i+1,k)\in O\left(i,j\right)}\left\{ c_{0}\left(S\right)-c_{0}\left(S\left[\left(i+1,k\right)\right)\right]\right\} \nonumber \\
 & - & \frac{\beta}{b^{K-1}}\left|S\left(K\right)\right|c_{0}\left(S\right)\label{eq:zeroth_order}
\end{eqnarray}

We would like to extract the lowest power of $\frac{1}{b}$ with a
non-zero coefficient. This is determined by the sites in $S$ that
are closest to the root of the tree (i.e. site $\left(1,1\right)$).
We must consider four different scenarios (note that $S$ can have
any structure further away from the root)
\begin{itemize}
\item Scenario 1; $S$ contains $\left(1,1\right)$ so that $\left(1,1\right)$
is the site closest to the root,
\item Scenario 2; $S$ contains a single bulk site that is closest to the
root,
\item Scenario 3; $S$ contains multiple bulk sites that are closest to
the root, and
\item Scenario 4; $S$ only contains sites at level $K$.
\end{itemize}
In scenario 1 the only terms that can contribute at zeroth order are
\begin{equation}
0=\alpha\left\{ c_{0}\left(S\left[\overline{\left(1,1\right)}\right]\right)-c_{0}\left(S\right)\right\} -\lim_{b\rightarrow\infty}\frac{1}{b}\sum_{(2,j)\in O\left(1,1\right)}\left\{ c_{0}\left(S\right)-c_{0}\left(S\left[\left(2,j\right)\right]\right)\right\} 
\end{equation}

where the sum in the second term is over the level $2$ sites not
in $S$. The number of such sites is of order $b$ so the limit is
required to extract the contribution at zeroth order. Notice that
all structure in $S$ beyond level 2 has no effect on the correlation
function.

If we now substitute the mean field hypothesis (\ref{eq:mft_ansatz})
we get
\begin{equation}
0=\alpha\left(1-t_{1}\right)-t_{1}\left(1-t_{2}\right)\lim_{b\rightarrow\infty}\left(\frac{b-\left|S\left(2\right)\right|}{b}\right).
\end{equation}

Using the fact that $\left|S\right|$ is independent of $b$, we find
that the mean field hypothesis is exact in the limit $b\rightarrow\infty$
and we recover the first boundary mean field equation (\ref{eq:mft_alpha}).
Therefore (\ref{eq:mft_alpha}) is valid for any correlation function
that contains the first site in the limit of large $b$.

In scenario 2 we assume that $S$ has a single site at level $l$
that is closest to the root. Equivalently
\[
\left|S\left(k\right)\right|=0\textnormal{ for }k<l\textnormal{ and }\left|S\left(l\right)\right|=1
\]

with the structure of $S$ arbitrary for $k>l$. We will assume without
loss of generality that the site at level $l$ is $\left(l,1\right)$.
In this case (\ref{eq:zeroth_order}) becomes
\begin{equation}
0=\left\{ c_{0}\left(S\left[\left(l-1,1\right),\overline{\left(l,1\right)}\right]\right)-c_{0}\left(S\left[\left(l-1,1\right)\right]\right)\right\} -\lim_{b\rightarrow\infty}\frac{1}{b}\sum_{\left(l+1,j\right)\in O\left(l,1\right)}\left\{ c_{o}\left(S\right)-c_{0}\left(S\left[\left(l+1,j\right)\right]\right)\right\} 
\end{equation}

where the sum in the second term is over all downstream neighbours
of $\left(l,1\right)$ that are not in $S$. Again the correlation
function does not depend on the structure of $S$ beyond the first
two occupied levels of the tree. If we now substitute the mean field
hypothesis (\ref{eq:mft_ansatz}) we get
\begin{equation}
0=\left\{ t_{l-1}\prod_{k=l+1}^{K}t_{k}^{\left|S\left(k\right)\right|}-t_{l-1}\prod_{k=l}^{K}t_{k}^{\left|S\left(k\right)\right|}\right\} -\left\{ \prod_{k=l}^{K}t_{k}^{\left|S\left(k\right)\right|}-t_{l+1}\prod_{k=l}^{K}t_{k}^{\left|S\left(k\right)\right|}\right\} \lim_{b\rightarrow\infty}\left(\frac{b-\left|O\left(l,1\right)\right|}{b}\right).
\end{equation}

Taking the $b\rightarrow\infty$ limit we find that mean field theory
is exact and assuming that none of the $t_{k}$ are zero we recover
the bulk mean field equation (\ref{eq:mft_bulk}). We have therefore
shown that to zeroth order in $\frac{1}{b}$ any correlation function
with a single site nearest the root satisfies mean field theory exactly.
Scenario 3 is a straightforward extension of this analysis and shows
that this conclusion holds even if we allow multiple sites at level
$l$.

In scenario 4 we assume that $S$ only includes sites at level $K$.
In this case equation (\ref{eq:zeroth_order}) reduces to
\begin{equation}
0=\left\{ c_{0}\left(S\left[\left(K-1,1\right),\overline{\left(K,1\right)}\right]\right)-c_{0}\left(S\left[\left(K-1,1\right)\right]\right)\right\} -\beta c_{0}\left(S\right).
\end{equation}

Substituting the mean field hypothesis (\ref{eq:mft_ansatz}) satisfies
this equation and recovers the second boundary mean field theory equation
(\ref{eq:mft_beta}).

In summary, we find that mean field theory is exact when we consider
a fixed correlation function and let $b\rightarrow\infty$. We expect
that mean field theory will be an increasingly good approximation
as the branching number increases. This bears out the plausibility
argument given at the start of the section and it seems reasonable
to expect to be true on other types of lattices and for more complex
exclusion processes. That is, mean field theory becomes an increasingly
good approximation as the branching number or coordination number
is increased for any ``TASEP on a network'' model where loops are
not significant. 

As a step beyond mean field theory we would like to look at $\frac{1}{b}$
corrections to the the correlation function expansion in equation
(\ref{eq:large_b_expansion}). Unfortunately this has proved much
more difficult than we originally hoped. At zeroth order we found
that all correlation functions can be expressed in terms of a single
site function $t_{k}$ using equation (\ref{eq:mft_ansatz}). It might
be expected that to first order in the expansion in $\frac{1}{b}$
we would have an analogous result but with two site correlation functions
(or at least a limited set of short range correlation functions) replacing
the single site functions. Indeed we do find that an arbitrary correlation
function can be expressed in terms of a reduced set of correlation
functions, but the reduced set is not limited to short range correlation
functions and in fact includes correlation functions that extend across
all layers of the tree. We have been unable to find an analytic approach
to dealing with this complexity and have chosen instead to make a
first step by solving the two level tree in the next section. We address
larger trees using Monte-Carlo simulation in Section \ref{sec:Simulation-results}.

\section{\label{sec:Exact-solution-for}Analytical study of the two level
tree}

In this section we make a full analysis for the case $K=2$ or two
level tree. As we shall see this is not a trivial problem to solve
for arbitrary $b$. We derive recursion relations for the correlation
functions, then first show how to obtain a $1/b$ expansion solution
to these. Then for the case $\alpha=1$ we obtain the exact solution
of the recursion recursion relations as a finite sum. We show how
to write this solution in an integral representation that easily allows
the $1/b$ expansion to be recovered by the saddle point method.

\subsection{Recursion relations for the two level tree}

Our starting point is to use the symmetry of the lattice to simplify
the correlation function $c_{S}$. In steady state $c_{S}$ will be
invariant under the exchange of any sub-branches of the tree. In the
$K=2$ case this means that any $c_{S}$ that has $m$ sites at level
2 is identical. As a consequence we can define 
\begin{equation}
c_{S}=c\left(l,m\right)
\end{equation}
where $l=0\,\textrm{or}\,1$ indicates whether site $\left(1,1\right)$
is included in the correlation function and $m=0,1,\ldots,b$ is the
number of sites included at level 2.

In the case $K=2$ and $l=1$ the steady state equation (\ref{eq:bl_steadystate})
becomes 
\begin{equation}
0=\alpha c(0,m)-\left(1+\alpha+\frac{m}{b}\left(\beta-1\right)\right)c\left(1,m\right)+\left(1-\frac{m}{b}\right)c\left(1,m+1\right)\quad\mbox{for}\quad m\geq0\,.\label{eq:k2_c1m_recursion}
\end{equation}
In the case $l=0$ the steady state equation gives 
\begin{equation}
0=c\left(1,m-1\right)-c\left(1,m\right)-\beta c\left(0,m\right)\quad\mbox{for}\quad m\geq1\label{eq:k2_c0m_recursion}
\end{equation}
with boundary condition $c(0,0)=1$.

We may eliminate $c(0,m)$ from (\ref{eq:k2_c1m_recursion}, \ref{eq:k2_c0m_recursion})
to obtain 
\begin{equation}
0=\frac{\alpha}{\beta}c(1,m-1)-\left(\frac{\alpha+\beta+\alpha\beta}{\beta}+\frac{m}{b}\left(\beta-1\right)\right)c\left(1,m\right)+\left(1-\frac{m}{b}\right)c\left(1,m+1\right)\quad\mbox{for}\quad m\geq1\,.\label{k2recursion2}
\end{equation}
Finally, the boundary case $m=0$ of (\ref{eq:k2_c1m_recursion})
yields 
\begin{equation}
0=\alpha-(1+\alpha)c(1,0)+c(1,1)\;.\label{c1m0}
\end{equation}

\subsection{$1/b$ expansion for a two level tree\label{sub:-expansion-for}}

In the limit $b\rightarrow\infty$ we can solve these recursion relations
easily to obtain 
\begin{eqnarray}
c\left(0,m\right) & = & t_{2}^{m},\label{eq:k2_mean_field_0}\\
c\left(1,m\right) & = & t_{1}t_{2}^{m},\label{eq:k2_mean_field_1}
\end{eqnarray}
where $t_{1}$ and $t_{2}$ are the same as the solutions to the mean
field theory equations (\ref{eq:mft_alpha}), (\ref{eq:mft_bulk})
and (\ref{eq:mft_beta}). They are given by 
\begin{equation}
t_{1}=1-\frac{\alpha+\beta+\alpha\beta-\sqrt{\left(\alpha+\beta+\alpha\beta\right)^{2}-4\alpha\beta}}{2\alpha}\label{eq:t1_mft}
\end{equation}
\begin{equation}
t_{2}=\frac{\alpha+\beta+\alpha\beta-\sqrt{\left(\alpha+\beta+\alpha\beta\right)^{2}-4\alpha\beta}}{2\beta}\,.\label{eq:t2_mft}
\end{equation}

We can develop an expansion around these mean field results by expanding
$c\left(l,m\right)$ as a power series in $\frac{1}{b}$ of the form
\begin{equation}
c\left(l,m\right)=\sum_{r=0}^{\infty}c_{r}\left(l,m\right)\frac{1}{b^{r}}\label{eq:cnm_expansion}
\end{equation}
where $c_{0}\left(l,m\right)$ is given by the mean field results
(\ref{eq:k2_mean_field_1}) and (\ref{eq:k2_mean_field_1}). Substituting
this expansion into equation (\ref{eq:k2_c1m_recursion}) and equating
successive powers of $\frac{1}{b}$ gives a sequence of recursion
relations that may be solved to obtain the coefficients in equation
(\ref{eq:cnm_expansion}).

As an illustration we give order $\frac{1}{b}$ correction 
\begin{eqnarray}
c\left(1,m\right) & = & t_{1}t_{2}^{m}+\frac{1}{b}\,\frac{\alpha\beta\left(\beta-1+t_{2}\right)t_{2}^{m+1}}{2\left(\alpha+1-t_{2}\right)^{2}\left(\beta t_{2}^{2}-\alpha\right)^{2}}\Bigl\{\left(\alpha+1-t_{2}\right)m^{2}\label{eq:c10_b_expansion}\\
 &  & -\left(\alpha+\beta t_{2}^{2}\right)\left(\alpha+1-t_{2}\right)m-2\alpha t_{2}\Bigl\}+\mathcal{O}\left(\frac{1}{b^{2}}\right)\,.
\end{eqnarray}

\begin{figure}
\noindent \begin{centering}
\includegraphics[scale=0.8]{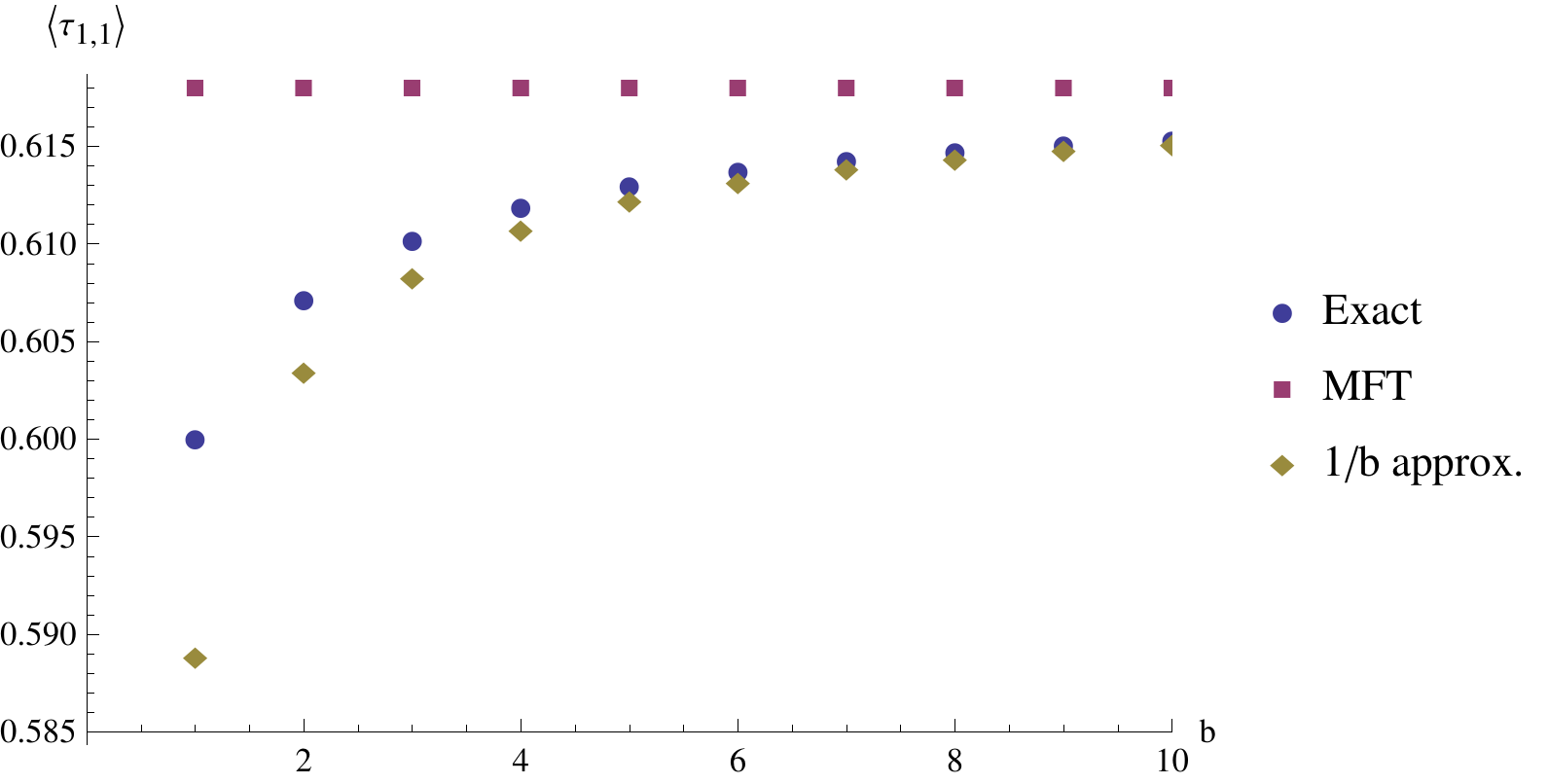} 
\par\end{centering}

\caption{\label{fig:Comparison-of-exact}Comparison of exact value of $\left\langle \tau_{1,1}\right\rangle $against
mean field and $\frac{1}{b}$ corrections for two level tree with
$\alpha=\beta=1$. The blue circles are the exact values from solving
the recursion relation for a given value of $b$, the dashed line
is the mean field estimate which is independent of $b$ and the red
squares are the result of expansion to $\mathcal{O}(\frac{1}{b})$
in equation (\ref{eq:c10_b_expansion}).}
\end{figure}

This can be used to approximate the average occupation number of the
first site $\left\langle \tau_{1,1}\right\rangle =c\left(1,0\right)$
for any value of $b$. This is plotted in figure \ref{fig:Comparison-of-exact}
and compared with the mean field theory estimate and exact results
obtained by solving the recursion relations exactly for small values
of $b$. The approximation is excellent for values of $b>5$.

This $1/b$ expansion approach is effective, but we have implicitly
made assumptions about the $b$ dependence of $c\left(l,m\right)$
in order to make the $1/b$ expansion of the recursion relation (\ref{eq:k2_c1m_recursion}).
We address this in the next section using an alternative approach
that confirms that the procedure does give correct results when $m\sim\mathcal{O}$$\left(1\right)$.

\subsection{Exact finite sum expressions for correlation functions of two level
tree}

In the case $\beta=1$ we can obtain a simple full exact solution
of the two level tree in the form of a finite sum. In the case $\beta=1$,
(\ref{k2recursion2}) becomes 
\begin{equation}
0=c(1,m-1)-\left(\frac{2\alpha+1}{\alpha}\right)c\left(1,m\right)+\frac{1}{\alpha}\left(1-\frac{m}{b}\right)c\left(1,m+1\right)\quad\mbox{for}\quad m\geq1\label{k2recursion3}
\end{equation}
We now change index to $n=b-m$ to express (\ref{eq:k2_c1m_recursion})
in the form of a recursion ``down'' from $b$ rather than ``up''
from $0$. Defining $D\left(n\right)=c\left(1,b-n\right)=c\left(1,m\right)$
equation (\ref{k2recursion3}) becomes 
\begin{equation}
D\left(n+1\right)=a\, D\left(n\right)-\frac{n}{b\alpha}D\left(n-1\right)\label{eq:d_recursion}
\end{equation}
for $n=0,1,\ldots b-1$, where 
\begin{equation}
a=\frac{2\alpha+1}{\alpha}.
\end{equation}

The first few terms in this recursion are easy to compute 
\begin{eqnarray*}
D(1) & = & aD(0)\\
D(2) & = & \left(a^{2}-\frac{1}{b\alpha}\right)D(0)\\
D(3) & = & \left(a^{3}-\frac{3a}{b\alpha}\right)D(0)\\
D(4) & = & \left(a^{4}-\frac{6a^{2}}{b\alpha}+\frac{3}{(b\alpha)^{2}}\right)D(0)\\
 & \vdots
\end{eqnarray*}
It can be proven by induction that the general solution for $D(n)$
is of the form 
\begin{equation}
D\left(n\right)=\sum_{r=0}^{\left\lfloor \frac{n}{2}\right\rfloor }\left(-1\right)^{r}\frac{a^{n-2r}}{\left(\alpha b\right)^{r}}\left(2r-1\right)!!\binom{n}{2r}\, D\left(0\right)
\end{equation}
where $\left(2r-1\right)!!=(2r-1)(2r-3)\cdots1$ with the convention
$\left(-1\right)!!=1$ and $\left\lfloor x\right\rfloor $ is the
floor function defined to be the largest integer less than or equal
to $x$.

Returning to the correlation function notation we have 
\begin{equation}
c\left(1,m\right)=c\left(1,b\right)S_{m}
\end{equation}
where 
\begin{equation}
S_{m}=\sum_{p=0}^{\left\lfloor \frac{b-m}{2}\right\rfloor }a^{b-m-2p}\left(\frac{-1}{\alpha b}\right)^{p}\binom{b-m}{2p}\left(2p-1\right)!!\,.\label{eq:s_sum}
\end{equation}
Finally to fix $c\left(1,b\right)$ we use the boundary condition
equation (\ref{c1m0}) to obtain 
\begin{equation}
c(1,b)=\frac{\alpha}{(1+\alpha)S_{0}-aS_{1}}\;.
\end{equation}
We may now write the exact solution for the correlation functions
as 
\begin{eqnarray}
c\left(1,m\right) & = & \frac{\alpha S_{m}}{\left(\alpha+1\right)S_{0}-S_{1}}\label{c1exact}\\
c\left(0,m\right) & = & \frac{\alpha\left(S_{m-1}-S_{m}\right)}{\left(\alpha+1\right)S_{0}-S_{1}}\;.\label{c0exact}
\end{eqnarray}
Thus equations (\ref{c1exact},\ref{c0exact}) give exact finite sum
expressions for all correlation functions for any $b$.

\subsection{Saddle-point expansion of exact solution}

We would like to obtain an asymptotic expansion in $1/b$ as in Section~\ref{sub:-expansion-for}
from expressions (\ref{c1exact},\ref{c0exact}). However the binomial
term involving $b$ in equation (\ref{eq:s_sum}) means that we do
not yet have the required form. To obtain an asymptotic form we first
introduce an integral representation of the double factorial (which
may be verified by integration by parts) 
\begin{equation}
\left(2p-1\right)!!=\frac{2^{p}}{\sqrt{\pi}}\intop_{-\infty}^{\infty}\mathrm{d}u\, u^{2p}e^{-u^{2}}\;.
\end{equation}

\begin{figure}
\noindent \begin{centering}
\includegraphics[scale=0.3]{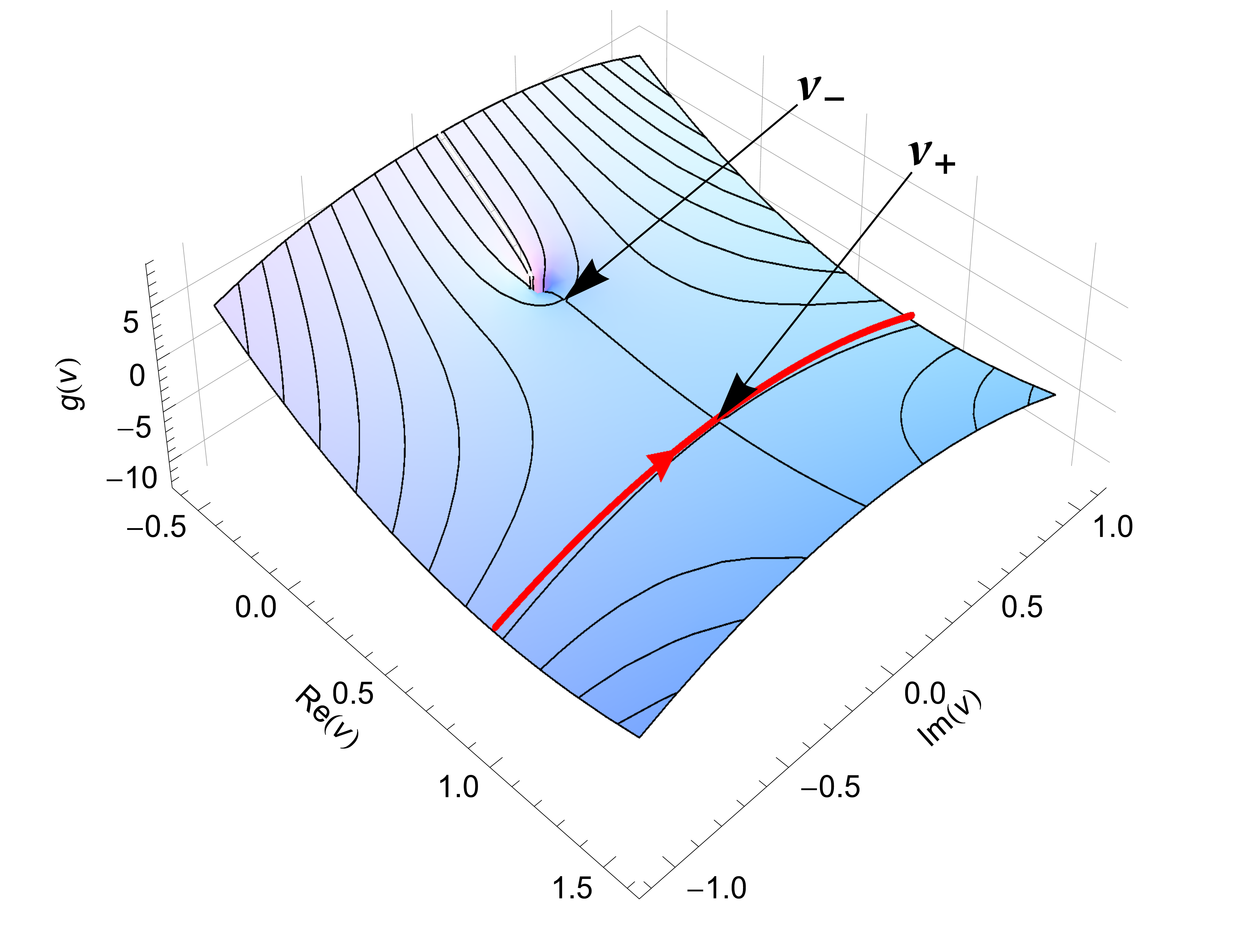}
\par\end{centering}

\caption{\label{fig:Contour-plot}Plot of the real part of $g\left(v\right)$
(see equation (\ref{eq:g_dfnn})) in the complex plane indicating
its two stationary points $v_{\pm}$. The physically relevant stationary
point is $v_{+}$.}
\end{figure}

Substituting into (\ref{eq:s_sum}) this gives 
\begin{eqnarray*}
S_{m} & = & \frac{a^{b-m}}{\sqrt{\pi}}\int_{-\infty}^{\infty}\mathrm{d}u\, e^{-u^{2}}\sum_{\begin{array}{c}
p=0\\
p\mathrm{\, even}
\end{array}}^{\infty}\left[\frac{iu\sqrt{2}}{a\sqrt{\alpha b}}\right]^{p}\binom{b-m}{p}\;.
\end{eqnarray*}
The series can be summed and with a change of variable $v=1+\frac{iu\sqrt{2}}{a\sqrt{\alpha b}}$
we can express $S_{m}$ as an integral in the form 
\begin{equation}
S_{m}=\frac{a^{b-m+1}\sqrt{\alpha b}}{i\sqrt{2\pi}}\int_{-i\infty}^{i\infty}\mathrm{d}v\,\frac{1}{v^{m}}\exp bg(v)\,.\label{eq:Sm_integral}
\end{equation}
where 
\begin{equation}
g(v)=\frac{a^{2}\alpha}{2}\left(v-1\right)^{2}+\ln v\,.\label{eq:g_dfnn}
\end{equation}
We can obtain an asymptotic expansion in $b$ using steepest descents,
but we need to make some assumptions about the behaviour of $m$.
We consider three regimes; 
\begin{description}
\item [{(i)}] Assume $m\sim\mathcal{O}\left(1\right)$. This is the case
of most practical interest; where the correlation function has a fixed
number (of order 1) of sites. It corresponds to the assumptions we
made in Section \ref{sub:Expansion-of-correlation} to show that MFT
is exact in the $b\rightarrow\infty$ limit. It also enables the computation
of density and current. 
\item [{(ii)}] Assume $\frac{m}{b}=d$ where $d\sim\mathcal{O}\left(1\right)$
and $0<d<1$. This corresponds to correlation functions with a finite
proportion of the total number of sites. 
\item [{(iii)}] Assume $b-m=n$ where $n\sim\mathcal{O}\left(1\right)$.
This corresponds to correlation functions containing nearly all sites. 
\end{description}
We address regime (i) first and take $b\gg1$ and $m\sim\mathcal{O}\left(1\right)$.
We can use steepest descents to obtain an asymptotic expansion in
$b$ from equation (\ref{eq:Sm_integral}). In this regime the pre-factor
$1/v^{m}$ is independent of $b$ and the stationary points for the
steepest descents are determined by $g\left(v\right)$ in equation
(\ref{eq:g_dfnn}). This has stationary points 
\begin{equation}
v_{\pm}=\frac{1}{2}\left\{ 1\pm\left(1-\frac{4}{\alpha a^{2}}\right)^{\nicefrac{1}{2}}\right\} \label{eq:v_plus_minus}
\end{equation}
and they are shown in figure \ref{fig:Contour-plot}. We see that
the $\ln v$ term in equation (\ref{eq:g_dfnn}) leads to a branch
cut along the negative real axis and a logarithmic singularity at
the origin. In order to use the saddle point approximation we deform
the contour in equation (\ref{eq:Sm_integral}) away from the imaginary
axis to run through one of the stationary points. This will not succeed
for the $v_{-}$ stationary point because the path of steepest descents
runs along the real axis towards the origin and there is no return
path through low values of the integrand. With $v_{+}$ the path of
steepest descent is parallel to the imaginary axis and the contour
can easily be chosen so that only points close to the saddle point
contribute to the integral.

Using this contour we find that the zeroth order term in the asymptotic
expansion is given by
\begin{equation}
c\left(1,m\right)\sim\frac{\alpha}{1+\alpha-\frac{1}{av_{+}}}\left(av_{+}\right)^{-m}\label{eq:zeroth_i}
\end{equation}
If we compare equation (\ref{eq:t2_mft}) with equation (\ref{eq:v_plus_minus})
for $\beta=1$ we see that $t_{2}=1/(av_{+})$ and with a little algebra
we see that equation (\ref{eq:zeroth_i}) is equivalent to the MFT
result equation (\ref{eq:k2_mean_field_1}). In general the saddle
point is equivalent to MFT. This is in agreement with the results
of Section \ref{sub:Expansion-of-correlation} where we showed that
the MFT results become exact if we take a fixed correlation function
and then take the large $b$ limit. Further note that the same mean
field theory and saddle point hold for $m\ll b$ i.e. $m=o\left(b\right)$.

We can extend the asymptotic expansion to higher order. One can compute
the correction to the saddle point and one recovers the second term
in the expansion (\ref{eq:c10_b_expansion}). Similarly by computing
higher order corrections to the saddle point we can in principle compute
the full asymptotic expansion. For example with $\alpha=1$ and $m=0$
we obtain
\begin{equation}
c\left(1,0\right)=0.618034-\frac{0.0291796}{b}+\frac{0.0238699}{b^{2}}-\frac{0.032957}{b^{3}}+O\left(\left(\frac{1}{b}\right)^{4}\right).\label{eq:asym_3rd_order}
\end{equation}

\begin{figure}
\noindent \begin{centering}
\includegraphics[scale=0.8]{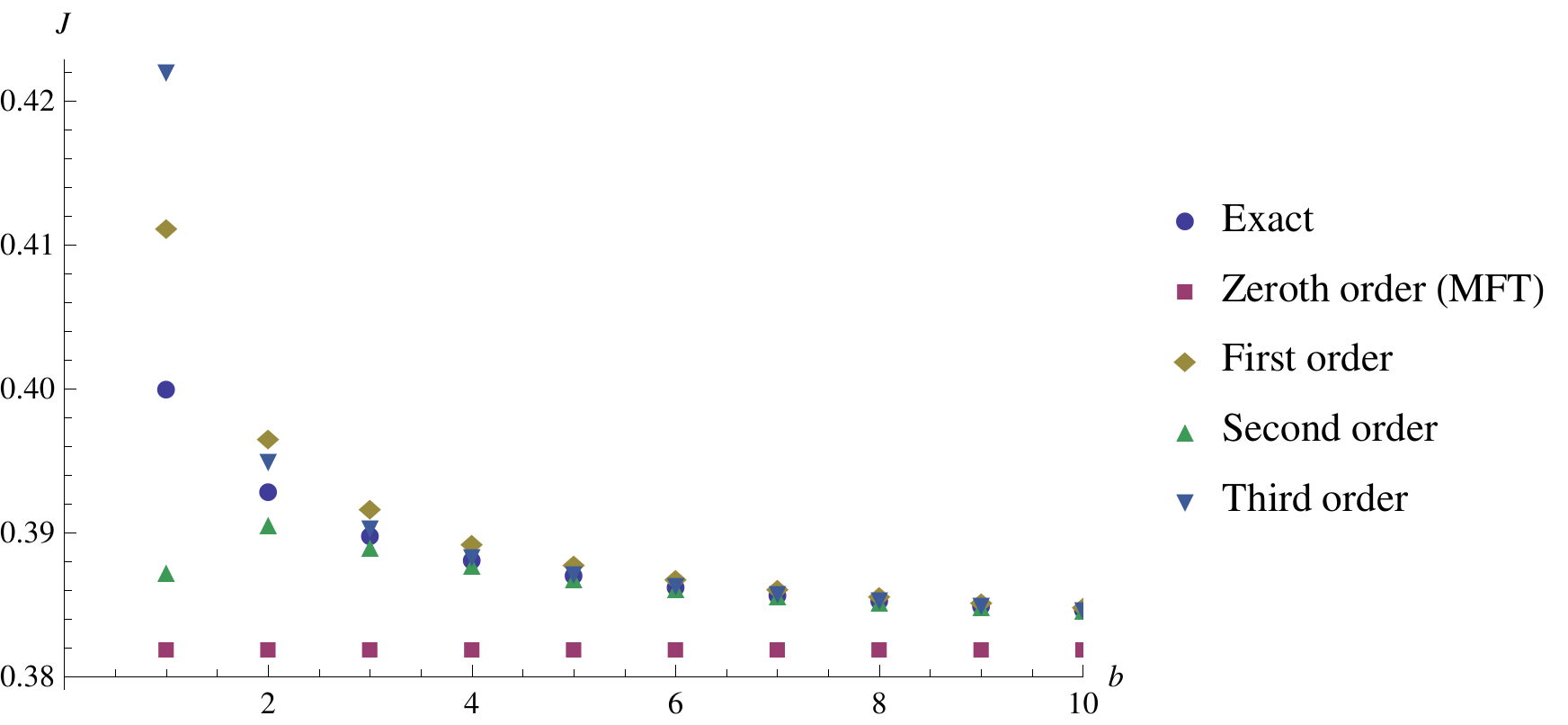}
\par\end{centering}

\caption{\label{fig:Current_k_2}The steady-state current $J$ versus $b$
for a two level tree ($K=2$) for $\alpha=\beta=1$. The exact results
are obtained by numerical solution of the steady-state equation (\ref{eq:bl_steadystate}).
The approximate values are obtained from the asymptotic expansion
of equation (\ref{eq:Sm_integral}) about $v_{+}$. The zeroth order
term is equivalent to MFT. The first order result includes only the
order $1/b$ term. Including second and third order terms in $1/b$
gives improved accuracy for small values of $b$, but provide little
improvement above $b=5$. }

\end{figure}

This result agrees with the first order correction given in equation
(\ref{eq:c10_b_expansion}). This result is used in figure \ref{fig:Current_k_2}
to plot the steady-state current using the relationship $J=\alpha\left(1-c\left(1,0\right)\right)$.
With the exception of $b=1$ additional terms improve the approximation,
but provide little improvement over the first order correction above
$b=5$.

We now turn to the other two regimes for $m.$ In regime (ii) with
$d=b/m$ we write equation (\ref{eq:Sm_integral}) and equation (\ref{eq:g_dfnn})
as
\begin{equation}
S_{m}=\frac{a^{b-m+1}\sqrt{\alpha b}}{i\sqrt{2\pi}}\int_{-i\infty}^{i\infty}\mathrm{d}v\,\exp bg_{d}(v).\label{eq:Sm_integral-c}
\end{equation}

and
\begin{equation}
g_{d}(v)=\frac{a^{2}\alpha}{2}\left(v-1\right)^{2}+\left(1-d\right)\ln v.\label{eq:gc_dfnn}
\end{equation}

The steepest descents contour now passes through one of the stationary
points of $g_{d}\left(v\right)$ which are given by 
\begin{equation}
v_{d}^{\pm}=\frac{1}{2}\left\{ 1\pm\left(1-\frac{4\left(1-d\right)}{\alpha a^{2}}\right)^{\nicefrac{1}{2}}\right\} .\label{eq:vc_plus_minus}
\end{equation}

Following a similar argument to regime (i), the contour of integration
runs parallel to the imaginary axis and passes through $v_{d}^{+}$ and
an asymptotic expansion can be obtained using steepest descents. The zeroth
order term is
\begin{equation}
c\left(1,m\right)=\frac{\alpha a^{-bd}}{1+\alpha-\frac{1}{av_{+}}}\sqrt{\frac{g^{''}\left(v_{+}\right)}{g_{d}^{''}\left(v_{d}^{+}\right)}}\exp\left\{ b\left(g_{d}\left(v_{d}^{+}\right)-g\left(v^{+}\right)\right)\right\} .\label{eq:zeroth_ii}
\end{equation}

Finally, in regime (iii), the same type of analysis can be repeated
to obtain a slightly simpler form
\begin{equation}
c\left(1,m\right)=\frac{\alpha a^{n-b-1}}{1+\alpha-\frac{1}{av_{+}}}\sqrt{\frac{g^{''}\left(v_{+}\right)}{\alpha}}\exp\left\{ -bg\left(v^{+}\right)\right\} .\label{eq:zeroth_iii}
\end{equation}

\begin{figure}
\noindent \begin{centering}
\includegraphics[scale=0.8]{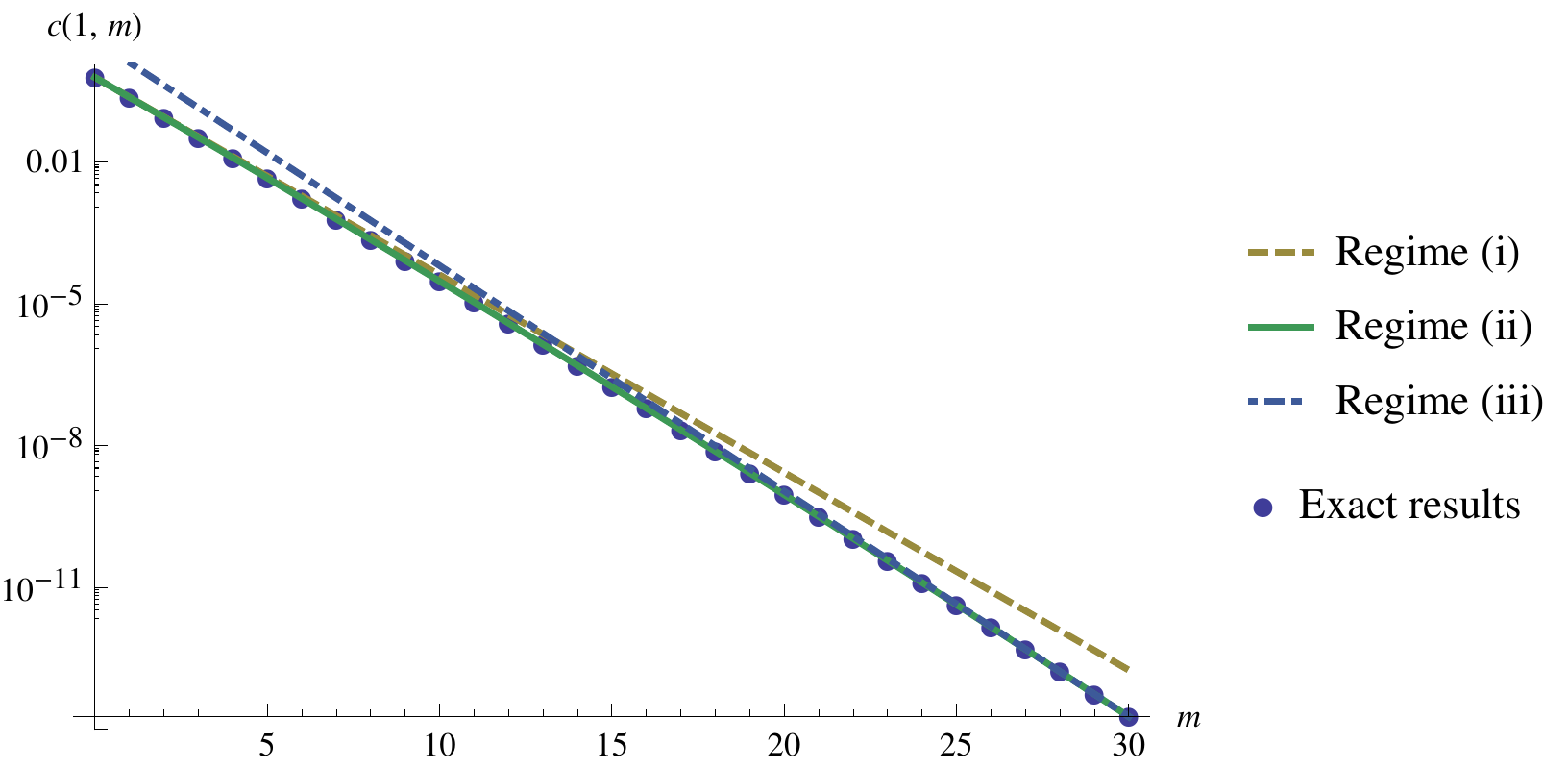}
\par\end{centering}

\caption{\label{fig:Comparison-of-the}Comparison of the three regimes of the
asymptotic expansion for the correlation function $c\left(1,m\right)$
with $b=30$ and $\alpha=\beta=1$. The exact results are from equation
(\ref{c1exact}), regime (i) is given by equation (\ref{eq:zeroth_i}),
regime (ii) is given by equation (\ref{eq:zeroth_ii}) and regime
(iii) is given by equation (\ref{eq:zeroth_iii}).}

\end{figure}

The $m$ dependence is a simple exponential through the parameter
$n$. In figure \ref{fig:Comparison-of-the} we plot the three zeroth
order asymptotic expansions for $c\left(1,m\right)$ versus the exact values. We see that in regime (i) (small $m$) equation (\ref{eq:zeroth_i}) provides a good approximation to the exact results and in regime (iii) ($m$ close to $b$)  equation (\ref{eq:zeroth_iii}) provides a good approximation to the exact results. In addition, we see that equation (\ref{eq:zeroth_ii}) actually gives an excellent fit in all three regimes. This is not surprising since  equation (\ref{eq:zeroth_ii}) reduces to equation (\ref{eq:zeroth_i}) in  regime (i) and  equation (\ref{eq:zeroth_iii}) in regime (iii).

\section{\label{sec:Simulation-results}Simulation results}

In order to investigate trees with $K>2$ with finite $b$ we used
Monte Carlo simulation. We simulated the system using a variant of
the waiting time algorithm \cite{gillespie_exact_1977} optimized
for best performance for our specific model. The state of the system
was described by a sequence $\{\tau_{k,i}\}$ of occupation numbers
(0 or 1), and a vector $\{n_{k}\}$ of the numbers $n_{k}$ of particles
in level $k$ that can jump to (free) sites in level $k+1$. Although
$\{\tau_{k,i}\}$ would suffice to fully define a microstate, the
use of the redundant vector $\{n_{k}\}$ improved the speed of the
algorithm. Namely, in each time step we calculated total rates $\{r_{k}=n_{k}/b^{k}\}$
of hopping from level $k$ to $k+1$ based on $n_{k}$, and then used
these to choose the level from which the hop would be attempted. The
selection was made by scanning through the list of all $K+1$ rates
(including the hop into the root with rate $\alpha(1-\tau_{1,1})$,
and the hop out the lattice with rate $\beta n_{K}/b^{K-1}$), and
since there were always not more than $K\leq20$ levels, this approach
was much faster than if we had just tried to pick up the departure
site directly out of $1+b+b^{2}+\dots+b^{K-1}$ possibilities. After
selecting the level $k$, the program chose an occupied site $i$
from this level and one of its empty neighbours $j$ from the next
level (by trial and error), and updated the variables $\{\tau_{k,i}\}$
and $\{n_{k}\}$, i.e., the particle jumped from $(k,i)$ to $(k+1,j)$.
Finally, the program increased the time by an exponentially-distributed
random variable $\Delta t$ with mean $1/R_{{\rm tot}}$, where $R_{{\rm tot}}$
was the sum of all rates.

Before any data were collected, the program performed a ``thermalization
run'' for 10\% of the total simulation time. Then, the program accumulated
the histogram of the density profile and the current every $N/2$
steps, weighting each of them by $\Delta t$. Each simulation was
repeated 10 times, starting each time from a different seed for the
random number generator to ensure statistical independence, and errors
of $\left<\tau_{k,i}\right>$ and $\left<J\right>$ were estimated
as standard errors.

To explore the behaviour of the TASEP on a tree for $K>2$ we have
used Monte Carlo simulations for system sizes up to 6 million sites.
The exponential growth in the system size with $K$, the number of
layers in the tree, limits the depth of tree we can easily simulate.
However, we present results for systems up to depth $K=20$ with $b=2$
and 1,048,575 sites and also up to branching ratio $b=50$ with $K=5$
and 6,377,551 sites. 

Our main objectives with the simulations are: to see how well MFT approximates systems with
finite $b$ and to see to what extent the behaviour of systems with
$b\geq2$ differs from the exact solution for the one-dimensional
($b=1$) case. We have focused on four representative points on the
phase diagram of figure \ref{fig:Phase-diagram}: $\alpha=0.25$ and
$\beta=1.0$ in the low density phase, $\alpha=1.0$ and $\beta=1.0$
in the maximal current phase, $\alpha=0.25$ and $\beta=0.25$ on
the boundary between the low density phase and the high density phase
and $\alpha=1.0$ and $\beta=0.25$ in the high density phase.

\begin{figure}[H]
\includegraphics[scale=0.75]{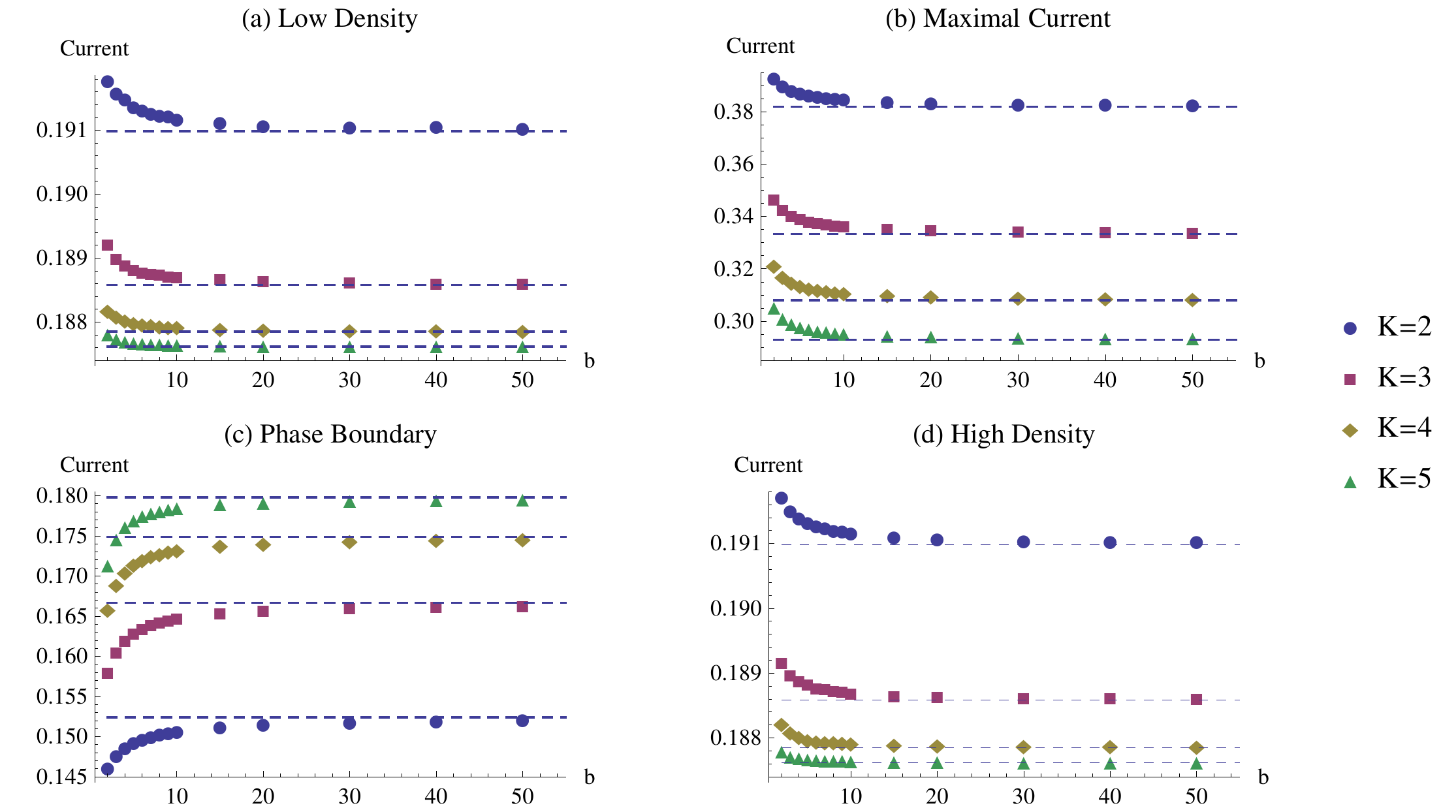}

\caption{\label{fig:current_vs_b}Plots for the TASEP on a tree of current
versus b for $K=2,3,4,5$ using the simulation results. The dashed
lines are the mean field theory results for the relevant values of
K. Details of the plots are; (a) is in the low density phase with
$\alpha=0.25$ and $\beta=1.0$. (b) is in the maximal current phase
with $\alpha=1.0$ and $\beta=1.0$. (c) is on the phase boundary
between low density and high density phases with $\alpha=0.25$ and
$\beta=0.25$. (d) is in the high density phase with $\alpha=1.0$
and $\beta=0.25$. }
\end{figure}

One interesting observation is that for a given value of $\alpha$
and $\beta$ on the phase diagram and a given value of $K$ the current
is a monotonically decreasing function of $b$ if $\alpha+\beta>1$
(see for example \ref{fig:current_vs_b}(a),(b) and (d)) and a monotonically
increasing function of $b$ if $\alpha+\beta<1$ (see for example
\ref{fig:current_vs_b}(c)). It is fairly easy to confirm analytically
that the $b\rightarrow\infty$ (MFT) current and the $b=1$ (one-dimensional)
current satisfy this, but we have been unable to find an analytic
proof of the more general result. We also note that this is consistent
with the result of equation (\ref{eq:mf_solution}) where we showed
that on the line $\alpha+\beta=1$ MFT is exact for any value of $b$
and therefore the current is independent of $b$.

In figure \ref{fig:current_vs_b} we show the relationship between
current and $b$ for different values of $K$. In each case the current
converges towards the MFT current as $b$ increases. However, on the
phase boundary, figure \ref{fig:current_vs_b}(c), this convergence
is much slower.

\begin{figure}[H]
\noindent \begin{raggedright}
\includegraphics[scale=0.75]{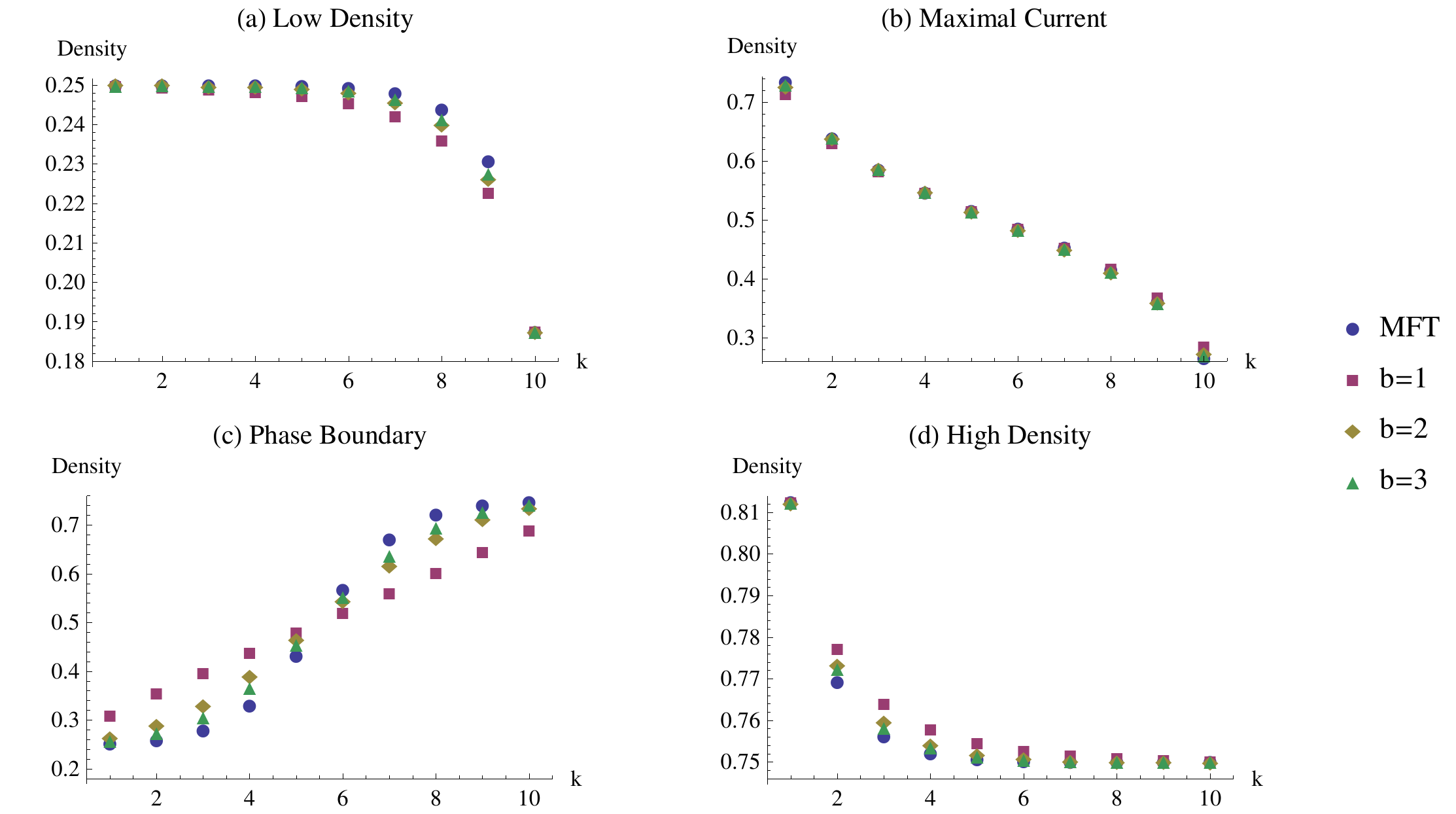}
\par\end{raggedright}

\caption{\label{fig:Density-profiles}Density profiles for TASEP on a tree
for $K=10$. The $b=2$ points are from simulation. These are compared
with the $b=1$ results from the exact solution of the one-dimensional
TASEP and the mean field theory which is exact in the limit $b\rightarrow\infty$.
Details of the plots are: (a) is in the low density phase with $\alpha=0.25$
and $\beta=1.0$; (b) is in the maximal current phase with $\alpha=1.0$
and $\beta=1.0$; (c) is on the phase boundary between low density
and high density phases with $\alpha=0.25$ and $\beta=0.25$; (d)
is in the high density phase with $\alpha=1.0$ and $\beta=0.25$. }

\end{figure}

In figure \ref{fig:Density-profiles} we show the relationship between
$k$, the level in the tree, and the average density of sites at that
level for each of the four points on the phase diagram. Simulation
results are shown for $b=2$ and $b=3$. These are compared against
the exact results for the $b=1$ exact solution and the $b\rightarrow\infty$
mean field solution.

\begin{figure}[H]
\includegraphics[scale=0.75]{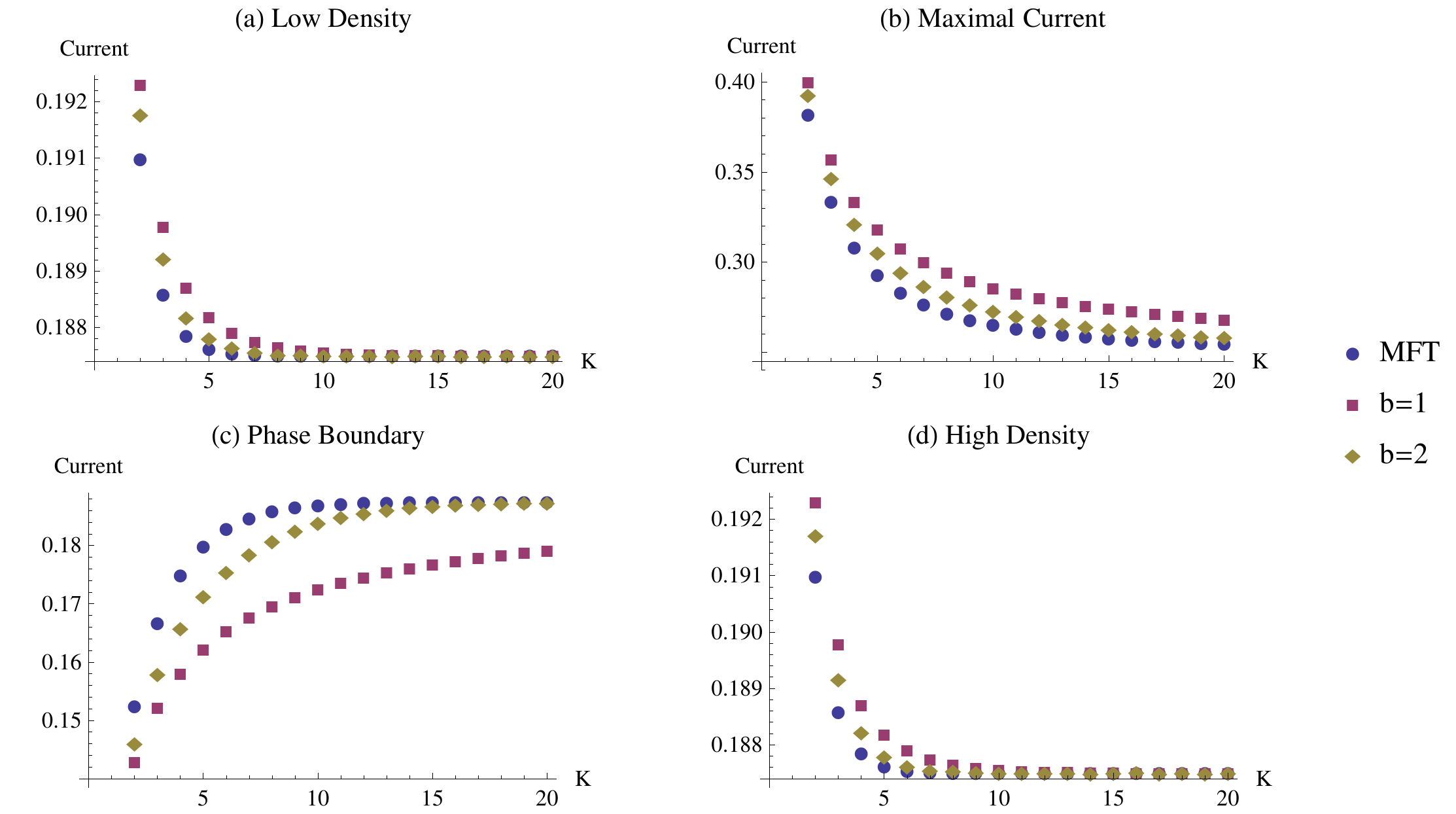}

\caption{\label{fig:Current_vs_K}Plots for the TASEP on a tree of current
versus K for $b=1$ using the exact 1D results, $b=2$ from the simulation
results and for mean field theory. Details of the plots are; (a) is
in the low density phase with $\alpha=0.25$ and $\beta=1.0$. (b)
is in the maximal current phase with $\alpha=1.0$ and $\beta=1.0$.
(c) is on the phase boundary between low density and high density
phases with $\alpha=0.25$ and $\beta=0.25$. (d) is in the high density
phase with $\alpha=1.0$ and $\beta=0.25$. }
\end{figure}

In figure \ref{fig:Current_vs_K} we show the relationship between
current and $K$ for $b=2$ at the four points on the phase diagram.
In each case these are compared with the corresponding mean field
result and with the exact results for one dimension which is equivalent
to $b=1$. In every case we find that the $b=2$ current lies between
the MFT and $b=1$ current and that all three converge towards the
same value for large $K$. 

In summary, the simulations demonstrate that values of $b$ greater
than one interpolate between the behaviour of the one-dimensional
model and MFT. The convergence towards mean field behaviour is rapid
with even $b=2$ being closer to mean field than to the $b=1$ behaviour
in most cases. We have looked at estimating critical exponents for
finite size behaviour and found the results inconclusive because of
the relatively small values of K of the systems simulated.

\section{Conclusion }

In this paper we have investigated a generalisation of the one-dimensional
TASEP to a tree lattice where the aggregate hopping rate from level
to level remains constant. We have shown with this choice for hopping
rates the TASEP on a tree shows a similar rich behaviour to the one-dimensional
case. The mean field theory for the tree models is identical to the
one-dimensional and predicts the correct phase diagram (which we confirm
by simulation). We have shown that mean field theory becomes exact
in the limiting case of large branching ratio or coordination number. 

We have presented an exact solution for the two level tree ($K=2$)
that allows the computation of any correlation function. Evaluating
the integral representation of the solution by the saddle point method
confirms the validity of MFT for an $n$-point correlation function
when $n\ll b$ and we take $b\rightarrow\infty$. We also computed
the steady-state current as a function of branching ratio $b$. The
form of the solution is fairly complex for $K=2$ and we were not
able to generalise it to higher values of $K$. 

Our simulation results indicate that the convergence to the large
branching ratio limit is quite rapid. We hope that the large branching
ratio limit will be useful in a broader set of situations than we
have dealt with here and we hope it may prove a useful limit for exploring
more complex aspects of the TASEP, such as disorder, which are not
tractable in the one-dimensional case. 

With regard to applications of the model, it has the appealing characteristic
that the ``total hopping capacity'' is the same at each level of
the tree. Any natural process that generates a tree like distribution
network by successively dividing capacity will approximate this characteristic.
As mentioned in the introduction the optimal design of these systems
has been the main focus of research in this area. However, the model
we analyse here attempts to answer a different question; what steady-state
current is achieved for specific values of the source input rate,
transport rate in the distribution network and sink output rate. Our
results show that the steady-state current has a non-trivial dependence
on these parameters. Interestingly, we provide a full analytic solution
for the two level tree (or star network or explosion network) that
is one of the simplest optimal networks. It remains an open question
as to which application areas are best addressed using a TASEP type
model with our model for the hopping rates, but vehicular traffic
is a reasonable candidate. Where it is applicable it predicts that
the system can potentially experience a high density, low density
or maximal current phase depending on how the parameters $\alpha,\beta$
position the system on the phase diagram Figure \ref{fig:Phase-diagram}.

As a next step it would be nice to develop an approximation for the
general tree that provides corrections to mean field theory and introduces
$b$ dependence. It would also be valuable to look at other non-equilibrium
processes on the tree.

\section*{Acknowledgments}
We would like to acknowledge a careful review of the paper by Richard Blythe. In addition, B.W. acknowledges the support of a Leverhulme Trust Early Career Fellowship and M.R.E. acknowledges the support of EPSRC under Programme grant number EP/J007404/1.

\appendix

\section{\label{sec:Equations-for-time}Equations for time evolution of a
general correlation function for the TASEP on a general lattice }

We can derive a very general result for the TASEP correlation functions
on a arbitrary network using the master equation.  In this appendix
we use the master equation to derive an equation for the time evolution
of the correlation function $c_{S}\left(t\right)$ defined in equation
(\ref{eq:crln_fncn_dfnn}) and we show that the time evolution of
$c_{S}\left(t\right)$ depends only the boundary sites of $S$ and
their nearest neighbours outside of $S$.

Taking the time derivative of $c_{S}$$\left(t\right)$ and substituting
for the time derivative of $P\left(\mathcal{C},t\right)$ using the
master equation (\ref{eq:master-simple}) gives
\begin{equation}
\frac{\partial c_{S}\left(t\right)}{\partial t}=\sum_{\mathcal{C^{\prime}}}P\left(\mathcal{C^{\prime}},t\right)\sum_{\mathcal{C}}W\left(\mathcal{C^{\prime}},\mathcal{C}\right)\left\{ \left(\prod_{i\in S}\tau_{i}\right)-\left(\prod_{i\in S}\tau_{i}^{\prime}\right)\right\} .\label{eq:appendix_2}
\end{equation}

The term in curly brackets has the effect of restricting the sums
on $\mathcal{C}$ and $\mathcal{C^{\prime}}$ to configurations where
all $\tau$ in $S$ are one or all $\tau^{\prime}$ in $S$ are one
but not both.

We now focus on a TASEP on a general directed graph where particles
can hop from site $i$ to site $j$ if there is a link from $i$ to
$j$ and we specify the hopping rate as $h_{ij}$. In addition we
identify a subset of ``entry'' sites where particles are introduced
at a site dependent rate $\alpha_{i}$ and a subset of ``exit''
sites where particles are removed at rate $\beta_{j}$. This gives
a transition matrix
\[
W\left(\mathcal{C^{\prime}},\mathcal{C}\right)=\begin{cases}
\alpha_{i} & \text{if }\mathcal{C^{\prime}},\mathcal{\, C}\:\text{are identical except at one entry site where }\tau_{i}^{\prime}=0,\tau_{i}=1,\\
h_{ij} & \text{if }\mathcal{C^{\prime}},\mathcal{\, C}\:\text{are identical except at two connected sites where }\tau_{i}^{\prime}=1,\tau_{i}=0,\tau_{j}^{\prime}=0,\tau_{j}=1,\\
\beta_{i} & \text{if }\mathcal{C^{\prime}},\mathcal{\, C}\:\text{are identical except at one exit site where }\tau_{i}^{\prime}=1,\tau_{i}=0,\\
0 & \mathrm{otherwise.}
\end{cases}
\]

We can express this transition matrix as;
\begin{eqnarray*}
W\left(\mathcal{C^{\prime}},\mathcal{C}\right) & = & \sum_{i\in S_{entry}^{all}}\alpha_{i}\left(1-\tau_{i}^{\prime}\right)\tau_{i}\prod_{j\in S^{all}\left[\overline{i}\right]}\delta\left[\tau_{j}^{\prime}=\tau_{j}\right]\\
 & + & \sum_{\left(i,j\right)\in E}h_{ij}\tau_{i}^{\prime}\left(1-\tau_{i}\right)\left(1-\tau_{j}^{\prime}\right)\tau_{j}\prod_{k\in S^{all}\left[\overline{i},\overline{j}\right]}\delta\left[\tau_{k}^{\prime}=\tau_{k}\right]\\
 & + & \sum_{i\in S_{exit}^{all}}\beta_{i}\tau_{i}^{\prime}\left(1-\tau_{i}\right)\prod_{j\in S^{all}\left[\overline{i}\right]}\delta\left[\tau_{j}^{\prime}=\tau_{j}\right]
\end{eqnarray*}

where the $\delta[\textrm{statement}]$ is a generalised Kronecker
delta; equal to unity if statement is true and zero otherwise. $E$
is the set of all directed edges,$S^{all}$ is the set of all sites
in the lattice and the subsets are defined as
\begin{eqnarray*}
S_{entry}^{all} & = & \text{subset of all entry sites,}\\
S_{exit}^{all} & = & \text{subset of all exit sites}\\
S^{all}\left[\overline{i},\overline{j},\ldots\right] & = & \text{subset of all sites excluding sites }i,j,\ldots
\end{eqnarray*}

If we substitute this expression for $W\left(\mathcal{C^{\prime}},\mathcal{C}\right)$
into equation (\ref{eq:appendix_2}) it has the effect of picking
out terms at the boundary of $S.$ Some algebra gives the general
result
\begin{eqnarray}
\frac{\partial c\left(S,t\right)}{\partial t} & = & \sum_{i\in S_{entry}}\alpha_{i}\left\{ c\left(S\left[\overline{i}\right],t\right)-c\left(S,t\right)\right\} \nonumber \\
 & + & \sum_{j\in S_{in}}\sum_{i\in I\left(j\right)}h_{ij}\left\{ c\left(S\left[i,\overline{j}\right],t\right)-c\left(S\left[i\right],t\right)\right\} \nonumber \\
 & - & \sum_{i\in S_{out}}\sum_{j\in O\left(i\right)}h_{ij}\left\{ c\left(S,t\right)-c\left(S\left[j\right],t\right)\right\} \nonumber \\
 & - & \sum_{i\in S_{exit}}\beta_{i}c\left(S,t\right)\label{eq:general_eqtn_of_mtn}
\end{eqnarray}

Where $S\left[\overline{i}\right]$ is the subset of sites obtained by subtracting the site $i$ from
$S$, $S\left[i\right]$ is obtained by adding site $i$ to
$S$ and $S\left[i,\overline{j}\right]$is obtained by adding
site $i$ and subtracting site $j$. The subsets used for the sums
in equation (\ref{eq:general_eqtn_of_mtn}) are defined in table (\ref{tab:Definition-of-boundary}). 

The physical meaning of equation (\ref{eq:general_eqtn_of_mtn}) is
clear if we take $S$ to contain a single bulk site; it is then a
simple statement of conservation of particles. We can extend this
to arbitrary $S$ by dividing all possible configurations into two
sets; those with all sites in $S$ occupied and those with at least
one vacant. The equation can then be obtained by considering the rate
of transitions between the two sets.

\section{\label{sec:Inductive-proof-that}Inductive proof that MFT is exact
on the line $\alpha+\beta=1$}

In this appendix we will prove that any set of sites $S$ will satisfy
equation (\ref{eq:line_of_mean_field}) which we reproduce here for
convenience
\begin{eqnarray}
0 & = & \delta\left[\left(1,1\right)\in S\right]+\sum_{\left(i,j\right)\in S_{in}}\frac{1}{b^{i-1}}-\sum_{\left(i,j\right)\in S_{out}}\frac{1}{b^{i}}\sum_{(i+1,k)\in O\left(i,j\right)}1-\sum_{i\in S_{exit}}1.\label{eq:line_of_mean_field-1}
\end{eqnarray}

If $S$ is the empty set then equation (\ref{eq:line_of_mean_field-1})
is clearly satisfied as all terms on the right hand side are zero.
If we now assume that it is true for an arbitrary $S$ then we need
to prove that it also holds for any $S^{\prime}$ constructed by adding
any single site to $S$. We can then assert the inductive step starting
with the empty set and adding sites to obtain any possible collection
of sites. When adding sites we need to consider eight possible scenarios
which modify the subsets of $S$ ($S_{in}$,$S_{out}$, $\left\{ \left(1,1\right)\right\} $,
$S_{exit}$) that appear in equation (\ref{eq:line_of_mean_field-1})
in different ways.
\begin{enumerate}
\item Add a site at $\left(1,1\right)$ that is disconnected from any site
in $S$. 
\item Add a site at $\left(1,1\right)$ that is connected to $m$ sites
in $S$ at level 2.
\item Add an exit site $\left(K,i\right)$ that is disconnected from any
site in $S$.
\item Add an exit site $\left(K,i\right)$ that is connected to $m$ sites
in $S$ at level $K-1$.
\item Add a bulk site $\left(k,i\right)$ that is disconnected from any
site in $S$.
\item Add a bulk site $\left(k,i\right)$ that is connected to $m$ sites
in $S$ at level $k+1$ but no sites at level $k-1$.
\item Add a bulk site $\left(k,i\right)$ that is connected to one site
in $S$ at level $k-1$ but no sites at level $k+1$.
\item Add a bulk site $\left(k,i\right)$ that is connected to one site
in $S$ at level $k-1$ and $m$ sites at level $k+1$.
\end{enumerate}
\begin{table}[H]
\noindent \centering{}%
\begin{tabular}{|c|c|c|c|c|c|}
\hline 
Scenario & First term & Second term & Third term & Fourth term & Total\tabularnewline
\hline 
\hline 
1 & $+1$ & $0$ & $-1$ & $0$ & $0$\tabularnewline
\hline 
2 & $+1$ & $-\frac{m}{b}$ & $-\frac{b-m}{b}$ & $0$ & $0$\tabularnewline
\hline 
3 & $0$ & $+\frac{1}{b^{K-1}}$ & $0$ & $-\frac{1}{b^{K-1}}$ & $0$\tabularnewline
\hline 
4 & $0$ & $0$ & $+\frac{1}{b^{K-1}}$ & $-\frac{1}{b^{K-1}}$ & $0$\tabularnewline
\hline 
5 & $0$ & $+\frac{1}{b^{k-1}}$ & $-\frac{1}{b^{k-1}}$ & $0$ & $0$\tabularnewline
\hline 
6 & $0$ & \selectlanguage{english}%
$+\frac{1}{b^{k-1}}-\frac{m}{b^{k}}$\selectlanguage{british}%
 & $-\frac{b-m}{b^{k}}$ & $0$ & $0$\tabularnewline
\hline 
7 & $0$ & $0$ & \selectlanguage{english}%
$+\frac{1}{b^{k-1}}-\frac{b}{b^{k}}$\selectlanguage{british}%
 & $0$ & $0$\tabularnewline
\hline 
8 & $0$ & $-\frac{m}{b^{k}}$ & \selectlanguage{english}%
$+\frac{1}{b^{k-1}}-\frac{b-m}{b^{k}}$\selectlanguage{british}%
 & $0$ & $0$\tabularnewline
\hline 
\end{tabular}

\caption{\label{tab:add_a_site}The change in each term on the right hand side
of equation (\ref{eq:line_of_mean_field-1}) from adding a single
site to $S$ using each of the possible scenarios described in the
text.}
\end{table}

In table \ref{tab:add_a_site} we evaluate the impact of each scenario
on the right hand side of equation (\ref{eq:line_of_mean_field-1}).
As we see the total change in the right hand side is zero in each
scenario, so that we have proven that if $S$ satisfies equation (\ref{eq:line_of_mean_field-1})
then so does $S^{'}$. Therefore, starting with the empty set we can
construct any possible set of sites and by induction it will also
satisfy (\ref{eq:line_of_mean_field-1}).

\bibliographystyle{unsrt}
\phantomsection\addcontentsline{toc}{section}{\refname}\bibliography{tasep_on_a_tree_v2}

\end{document}